\documentclass{aa}  

\usepackage{graphicx}
\usepackage{float}
\usepackage{txfonts}

\usepackage{hyperref}%
\hypersetup{colorlinks=True,citecolor=blue,linkcolor=blue}
\usepackage[nameinlink]{cleveref}

\begin{document}

   \title{A multi-band AGN-SFG classifier for extragalactic radio surveys using machine learning}


   \author{J. Karsten
          \inst{1}
          \and
          L. Wang\inst{1,2}
          \and
          B. Margalef-Bentabol\inst{2}
          \and
          P. N. Best\inst{3}
          \and
          R. Kondapally \inst{3}
          \and
          A. La Marca \inst{1,2}
          \and
          R. Morganti \inst{1,4}
          \and
          H.J.A. R\"ottgering \inst{5}
          \and
          M. Vaccari \inst{6,7,8}
          \and
          J. Sabater \inst{3, 9}
          }

   \institute{Kapteyn Astronomical Institute, University of Groningen, Groningen 9747, AD, The Netherlands
    \and
    SRON Netherlands Institute for Space Research, Landleven 12, 9747 AD, Groningen, The Netherlands
    \and
   Institute for Astronomy, University of Edinburgh, Royal Observatory, Blackford Hill, Edinburgh, EH9 3HJ, UK
    \and
   ASTRON, the Netherlands Institute for Radio Astronomy, Oude Hoogeveensedijk 4, 7991 PD Dwingeloo, The Netherlands
   \and
   Leiden Observatory, Leiden University, P.O.Box 9513, NL-2300 RA, Leiden, The Netherlands
   \and
   Inter-University Institute for Data Intensive Astronomy, Department of Astronomy, University of Cape Town, 7701 Rondebosch, Cape Town, South Africa
   \and
   Inter-University Institute for Data Intensive Astronomy, Department of Physics and Astronomy, University of the Western Cape, Robert Sobukwe Road, 7535 Bellville, Cape Town, South Africa
   \and
   INAF - Istituto di Radioastronomia, via Gobetti 101, 40129 Bologna, Italy
   \and
   UK Astronomy Technology Centre, Royal Observatory, Blackford Hill, Edinburgh, EH9 3HJ, UK
             }

   \date{Received XYZ; accepted XYZ}

 
  \abstract
   {Extragalactic radio continuum surveys play an increasingly more important role in galaxy evolution and cosmology studies. While radio galaxies and radio quasars dominate at the bright end, star-forming galaxies (SFGs) and radio-quiet Active Galactic Nuclei (AGNs) are more common at fainter flux densities.}
   {Our aim is to develop a machine learning classifier that can efficiently and reliably separate AGNs and SFGs in radio continuum surveys.
   }
   {We perform supervised classification of SFGs vs AGNs using the Light Gradient Boosting Machine (LGBM) on three LOFAR Deep Fields (Lockman Hole, Bo\"otes and ELAIS-N1), which benefit from a wide range of high-quality multi-wavelength data and classification labels derived from extensive spectral energy distribution (SED) analyses.
   }
   {Our trained model has a precision of 0.92$\pm$0.01 and a recall of 0.87$\pm$0.02 for SFGs. For AGNs, the model has slightly worse performance, with a precision of 0.87$\pm$0.02 and recall of 0.78$\pm$0.02. These results demonstrate that our trained model can successfully reproduce the classification labels derived from detailed SED analysis.  The model performance decreases towards higher redshifts, mainly due to smaller training sample sizes. To make the classifier more adaptable to other radio galaxy surveys, we also investigate how our classifier performs with a poorer multi-wavelength sampling of the SED. In particular, we find that the far-infrared (FIR) and radio bands are of great importance. We also find that higher S/N in some photometric bands leads to a significant boost in the model's performance. In addition to using the 150 MHz radio data, our model can also be used with 1.4 GHz radio data. Converting 1.4 GHz to 150 MHz radio data reduces performance by $\sim$4\% in precision and $\sim$3\% in recall. The final trained model is publicly available at \url{https://github.com/Jesper-Karsten/MBASC}}
   {}


   \maketitle
%

\section{Introduction}
Virtually all known massive galaxies host supermassive black holes (SMBHs) at their centres \citep{Kormendy_2013}. When such a black hole releases large amounts of energy by accreting gas rapidly, it can be observed as an Active Galactic Nucleus (AGN). AGNs are of great importance in studying galaxy evolution, as strong correlations exist between the SMBH mass and physical properties of the host galaxy, such as its velocity dispersion and bulge mass \citep{Ferrarese2000, Gebhardt2000, Kormendy_2013}. In addition, the cosmic black hole accretion history is similar to the cosmic star-formation history \citep{Kormendy_2013}. Theoretically, the energy released from AGNs can heat or expel the gas in the interstellar medium and quench star-formation activity in the host galaxies \citep[a mechanism known as AGN feedback;][]{Fabian2012, King2015}. This could explain why galaxies we see today are not as bright or massive as we might expect them to be from models and numerical simulations, which do not include AGN feedback \citep{Bower2006}.

Radio continuum surveys play a critical role in the detection of AGNs, particularly in finding the jet-mode AGNs. Observations in the radio can detect synchrotron radiation powered by the central SBMHs and/or recent star-formation activity. Early type galaxies normally emit synchrotron radiation at $<4\cdot 10^{20}$ W Hz$^{-1}$ at GHz radio frequencies from interstellar relativistic electrons \citep{Phillips1986, Sadler1989}. Radio galaxies, on the other hand, have radio GHz emission at $>10^{22}$ W Hz$^{-1}$ \citep{Sadler1989} due to relativistic jets. In the past, only the bright end of the radio sky could be probed, resulting in mostly detections of radio-loud galaxies. However, with more sensitive surveys, the faint end of the radio sky can also be probed. This results in modern radio surveys being able to probe not just radio galaxies but also radio-quiet AGNs (RQs) and star-forming galaxies (SFGs). Therefore, the need to efficiently and reliably classify different types of radio sources becomes increasingly more urgent.

Over the last few decades, many techniques have been developed to detect AGN activity in various parts of the electromagnetic spectrum. For example, ratios of certain emission lines are different for some AGNs from the typical O-stars in non-radiative sources. This means that their ratios of line fluxes can be analysed to find AGNs using so-called BPT diagrams \citep{Baldwin1981}. In the mid-infrared (MIR), photometric information can be used to find dust emission from the obscuring molecular gas and dust surrounding the black hole, which peaks at a rest frame of a ~few microns \citep[e.g.,][]{Stern2005, Donley2012}, which divide sources into AGNs and SFGs. X-ray data can detect emission from the accretion disk corona, which indicates AGN activity. Radio continuum emission can be used to locate the jets of AGNs. Lastly, spectral energy distribution (SED) analysis can be performed to detect the presence of an AGN, particularly if extensive multi-wavelength photometric information is available.

In terms of the classification scheme, AGNs can be classified into two categories based on their energetic output \citep{Heckman2014}. The first category includes
AGNs whose energetic output is mostly released via electromagnetic radiation produced by radiatively efficient
accretion of gas, which leads to the formation of an optically thick, geometrically thick accretion disk surrounding the SMBH \citep{Shakura1973}. This disk emits from the extreme ultraviolet (UV) through the visible in the electromagnetic spectrum \citep{peterson_1997, Osterbrock2006, krolik_1999}. Additionally, this disk is surrounded by a hot corona, which Compton-up-scatters photons into the X-ray band. The ionising radiation from the disk and the corona heats and ionises a portion of the gas clouds surrounding the AGN. This results in the production of emission lines in the UV,
optical, and near-infrared (NIR). Lastly, the accretion disk is also surrounded by a cloud of molecular gas. A portion of UV, visible and soft X-rays from the corona are absorbed by this dusty cloud and then emerge again as infrared emission. Traditionally these AGNs are known as Quasar-like. In this paper, we use the name `high-excitation' (due to their strong high-excitation emission lines) or `radiative-mode'.

The second category, known as jet-mode (or low-excitation)
AGNs constitute AGNs which produce little electromagnetic radiation compared to the first category. Their primary mode of energetic output is via kinetic energy transported in so-called jets (two-sided collimated beams of relativistic particles). It should be noted that a fraction of radiative-mode AGNs can also produce these jets. The geometrically-thin accretion disk mentioned for the other type of AGN is either absent or is replaced by a geometrically thick structure \citep{Quataert2001, Ho_2008}, which is consistent with the lower Eddington-scaled accretion rate. AGNs that have this excessive radio emission (as displayed by jets) are known as radio-loud. These can be identified by their aforementioned jets or by observing an excess radio emission compared to what is expected based on star-formation activity \citep{Gurkan2018, Smith2021}. The mechanism behind the generation of the jets is debated, but mechanisms involving rotating black holes and magnetic flux accretion are plausible \citep{Condon1984, Windhorst1985}. At low flux densities ($<$0.1 mJy), the source counts are dominated by RQs and SFGs. At increasing flux densities, the source counts quickly become dominated by radio-loud AGN above $\approx$ 1 mJy \citep{Padovani2015}.

These two binary criteria (radiative and radio-excess) described above can then be used to define four classes: SFGs (non-radiative and no radio-excess), RQs (radiative and no radio-excess), low-excitation radio galaxy (LERG) (non-radiative and radio-excess) and high-excitation radio galaxy (HERG) (radiative and radio-excess).

The main goal of this paper is to use supervised machine learning (ML) trained on classification labels obtained from previous SED analysis to create a fast and reliable method of classifying radio sources as AGNs or SFGs. The advantage of ML algorithms is that once they are trained, it is quick and easy to apply them to a new, similar dataset. In addition, ML classifications are always reproducible. We investigate supervised machine learning (ML) methods by using multi-wavelength photometry and photometric redshifts of radio sources detected in the first data release LOw-Frequency ARray (LOFAR) Two-metre Sky Survey (LoTSS) Deep Fields. The labels for these sources come from a detailed SED analysis with different SED fitting codes \citep{best2023lofar}.

This paper is organised as follows. In Sect. \ref{sec:data} we discuss the LOFAR radio data in the Deep Fields and the associated multi-wavelength photometric data on which the machine learning algorithm is trained. We also discuss how the separation between SFGs and AGNs was performed using SED analysis. In Sect. \ref{sec:classification} we describe the supervised ML algorithm adopted in this paper and the preprocessing, hyperparameters and metrics used. In Sect. \ref{sec:results} we present our results on the overall performance of the ML-based classifier, including a feature relevance study. In addition, we investigate how the performance of the classifier depends on factors such as sample size, SED sampling, and S/N of the various filters. Finally, in Sect. \ref{sec:conclusions}, we present our conclusions of this study as well as information on how to access our classifier for radio sources.

\section{Data}
\label{sec:data}
To apply supervised ML methods, labelled data are required. We use $\sim$80,000 radio sources in three LoTSS Deep Fields \citep{Tasse2021, Sabater2021, Duncan2021, Kondapally2021}: ELAIS-N1, Bo\"otes and Lockman Hole. These sources have been cross-matched to their multi-wavelength counterparts. \cite{best2023lofar} performed careful SED analysis using multiple fitting codes to classify the sources as SFG, RQ, HERG, or LERG. In this section, we present the key information regarding the LOFAR radio data and the associated multi-wavelength photometric data, as well as a brief summary of the SED-based classification process.

\subsection{The parent radio source catalogues and the associated multi-wavelength data}
\label{sec:multi-wavelength}
Radio observations in the three fields have been conducted using the LOFAR telescope \citep{Haarlem2013}. This instrument performs deep and wide radio observations of the sky thanks to its high sensitivity, high angular resolution, and wide field of view. The LoTSS Deep Fields are a deep survey which includes the European Large Area Infrared Space Observatory Survey Northern Field 1\citep[ELAIS-N1;][]{Oliver2000}), the Bo\"otes field \citep{jannuzi1999} and the Lockman Hole \citep{Lockman1986}. This survey has sufficient sky area to observe a full range of environments at wide redshift ranges, aiming to reach a noise level of 10-15 $\mu$Jy beam$^{-1}$ at 150 MHz. For the first data release, radio observations were taken with the High Band Antenna array (HBA) centred at roughly 150 MHz and are described by \cite{Tasse2021} for the Bo\"otes and Lockman Hole fields and by \cite{Sabater2021} for the ELAIS-N1 field. Source extraction is performed using Python Blob Detector and Source Finder \citep{Mohan2015}.  

Each of these three fields has extensive associated multi-wavelength data across a wide range of the electromagnetic spectrum \citep{Kondapally2021}. We summarise the data available in each field here. For a  detailed description of the multi-wavelength properties and cross-identifications of the radio sources, we refer the reader to \cite{Kondapally2021}.

FUV and NUV data come from data releases 6 and 7 of the Deep Imaging Survey (DIS) taken with the Galaxy Evolution Explorer (GALEX) space telescope \citep{Martin2005, Morrissey2007} for all three fields. The GALEX observations cover around 13.5 deg$^2$ in ELAIS-N1, 8 deg$^2$ in Bo\"otes, and also 8 deg$^2$ in Lockman Hole.

Observations in the u-band are taken from the Spitzer Adaptation of the Red-sequence Cluster Survey \citep[SpARCS;][]{Wilson2009, Muzzin2009} in ELAIS-N1 and the Lockman Hole covering $\sim$12 and $\sim$13 deg$^{2}$. For Bo\"otes the U-band data have been observed with the Large Binocular Telescope \citep[LBT;][]{Bian_2013}, which covers 9 deg$^2$.

In the optical, observations in the \textit{grizy} bands have been taken using the Panoramic Survey Telescope and Rapid Response System \citep[PanSTARRS;][]{Kaiser2010} in the Medium Deep Survey \citep[MDS;][]{Chambers2016} for ELAIS-N1. For Bo\"otes, the R- and I-band are taken as part of the NOAO Deep Wide Field Survey \citep[NDWFS;][]{jannuzi1999}, z-band data come from the zBo\"otes survey \citep{Cool_2007}, which covers the entire NDWFS field. Lastly, y-band data in Bo\"otes have been observed with the LBT covering the entire NDWFS field as well. The g, r and z band data have been taken by SpARCs in the Lockman Hole, while i-band has been observed within the Red Cluster Sequence Lensing Survey \citep[RCSLenS;][]{Hildebrandt2016}. MDS covers 8.05 deg$^2$ in ELAIS-N1, NDWFS covers 9.3 deg$^2$ in Bo\"otes, and RCSLenS covers 16.63 deg$^2$ in Lockman Hole.

NIR data in the J- and K-band come from the UK Infrared Deep Sky Survey Deep Extragalactic Survey (UKIDSS-DXS) Data Release 10 \citep{Lawrence2007} for ELAIS-N1 (covering 8.87 deg$^2$) and the Lockman Hole (covering 8.16 deg$^2$). These observations were done using the WFCAM instrument \citep{Casali2007} on the UK Infrared Telescope \citep[UKIRT;][]{Lawrence2007}. For Bo\"otes, the J, H and K-band data have been observed within the NOAO Extremely Wide-Field Infrared Imager \citep[NEWFIRM;][]{Whitaker2011, Gonzalez2010}, covering 8.5 deg$^2$.

The MIR data at 3.6, 4.5, 5.8, and 8.0 $\mu$m come from the Infrared Array Camera \citep[IRAC;][]{Fazio2004} on the Spitzer Space Telescope \citep{Werner2004} from the Spitzer Wide-area InfraRed Extragalactic (SWIRE) survey \citep{Lonsdale2003} for ELAIS-N1 (covering 9.32 deg$^2$) and the Lockman Hole (covering 10.95 deg$^2$). On the same telescope, the Spitzer Deep Wide Field Survey \citep{Ashby_2009} observed filters from 3.6 to 8.0 $\mu$m for Bo\"otes, covering approximately 10 deg$^2$.

The 24 $\mu$m data are taken using the Multi-band Imaging Photometer for Spitzer \citep[MIPS;][]{Rieke2004} covering all fields. 

Data at 100 $\mu$m and 160 $\mu$m were observed with the Photodetector Array Camera and Spectrometer \citep[PACS;][]{Griffin2010}. 250 $\mu$m, 350 $\mu$m and 500 $\mu$m were taken using the Spectral and Photometric Imaging Receiver \citep[SPIRE;][]{Poglitsch2010}. All data were taken within the Herschel Multi-tiered Extragalactic Survey \citep[HerMES;][]{Oliver2012} by the Herschel Space Observatory \citep{Pilbratt2010} covering all three fields. These data are part of the Herschel Extragalactic
Legacy Project \citep[HELP;][]{Shirley2021} with FIR deblending for the radio sources described by \cite{McCheyne2022}.

The above paragraphs do not describe the full extent of multi-wavelength data available in each field. We only include the data that we use. Some filters are only widely available in one field. We need consistent datasets over the field to use the ML algorithm on all three fields simultaneously, which gives us the maximum amount of data to train on. We, therefore, remove these filters. Often similar filters are available on different instruments (i.e., optical filters such as  g, r, i, z and y). A choice is then made for the filter with the most complete data. This is done to limit the number of missing values as more complete data means better performance of the model. Since not all fields have the same instruments and the same filters used to observe sources, some approximations have to be made. This means, in general, using similar filters/instruments to replace missing data (i.e., using the PanSTARRS i-band flux in ELAIS-N1 instead of NDWFS I-band, which is used in Bo\"otes). When there is no available equivalent band, the feature is then simply left empty. Non-detections and detections below 3$\sigma$ are left empty. Table \ref{table:wavelengths} shows the exact survey and corresponding depth for each field that is used for a specific feature. 
\begin{table*}[t]
\caption{Different filters and instruments used in each field. The 3$\sigma$ depths in AB magnitudes are provided for the FUV to IRAC ch4 bands in brackets. These depths were estimated using variances from empty 3" apertures. For the MIPS, PACS, and SPIRE bands, the limits at which fluxes can still be accurately deblended are given \citep{McCheyne2022}. For the radio data, the root mean
square (rms) sensitivity is given.}
\centering
\begin{tabular}{lll}
\hline
ELAIS-N1                    & Bo\"otes                  & Lockman Hole          \\ 
($\sim7.15$ deg$^2$)        & ($\sim10.73$ deg$^2$)     &  ($\sim9.5$ deg$^2$)  \\\hline
DIS FUV (26.3 [mag])        & DIS FUV (26.3 [mag])      & DIS FUV (26.3 [mag])                     \\
DIS NUV (26.7 [mag])        & DIS NUV (26.7 [mag])      & DIS NUV (26.7 [mag])                     \\
SpARCS u (25.4 [mag])       & SpARCS u (25.9 [mag])           & SpARCS u (25.5 [mag])       \\
PanSTARRS g (25.5 [mag])    & -                         & SpARCS g (25.8 [mag])       \\
PanSTARRS r (25.2 [mag])    & NDWFS R (25.2 [mag])      & SpARCS r (25.1 [mag])       \\
PanSTARRS i (25.0 [mag])    & NDWFS I (24.6 [mag])      & RCSLenS i (23.8 [mag])      \\
PanSTARRS z (24.6 [mag])    & zBoötes z (23.4 [mag])    & SpARCS z (23.5 [mag])       \\
PanSTARRS y (23.4 [mag])    & LBT y (23.4 [mag])        & -                     \\
UKIDSS-DXS J (23.2 [mag])   & NEWFIRM J (23.1 [mag])    & UKIDSS-DXS J (23.4 [mag])   \\
-                           & NEWFIRM H (22.5 [mag])    & -                     \\
UKIDSS-DXS K (22.7 [mag])   & NEWFIRM K (20.2 [mag])       & UKIDSS-DXS K (22.8 [mag])   \\
SWIRE ch1 (23.4 [mag])      & SDWFS ch1 (23.3 [mag])    & SWIRE ch1 (23.4 [mag])      \\
SWIRE ch2 (22.9 [mag])      & SDWFS ch2 (23.1 [mag])    & SWIRE ch2 (22.9 [mag])      \\
SWIRE ch3 (21.2 [mag])      & SDWFS ch3 (21.6 [mag])    & SWIRE ch3 (21.2 [mag])      \\
SWIRE ch4 (21.3 [mag])      & SDWFS ch4 (21.6 [mag])    & SWIRE ch4 (21.2 [mag])      \\
MIPS24 (20 [$\mu$Jy])       & MIPS24 (20 [$\mu$Jy])     & MIPS24 (20 [$\mu$Jy])         \\
PACS100 (12.5 [mJy])        & PACS100 (12.5 [mJy])      & PACS100 (12.5 [mJy])        \\
PACS160 (17.5 [mJy])        & PACS160 (17.5 [mJy])      & PACS160 (17.5 [mJy])        \\
SPIRE250 (4 [mJy])          & SPIRE250 (5 [mJy])        & SPIRE250 (4 [mJy])       \\
SPIRE350 (4 [mJy])          & SPIRE350 (5 [mJy])        & SPIRE350 (4 [mJy])       \\
SPIRE500 (6 [mJy])           & SPIRE500 (10 [mJy])       & SPIRE500 (6 [mJy])       \\
LoTSS (20 [$\mu$Jy])          & LoTSS (30 [$\mu$Jy])        & LoTSS (23 [$\mu$Jy])          \\\hline
\end{tabular}
\label{table:wavelengths}
\end{table*}

For a minority of sources, spectroscopic redshifts are available (1602, 4039, and 1466 sources in ELAIS-N1, Bo\"otes and Lockman, respectively); for the other sources, photometric redshifts are necessary. Photometric redshifts in all fields are obtained by using a combination of template fitting and ML methods \citep{Duncan2021}. \cite{Duncan2021} uses three template libraries: EAZY \citep{Brammer2008}, Extended Atlas of Empirical SEDs \citep{Brown2014} and Revised “XMM-COSMOS” Team templates \citep{Ananna2017}. Additionally, they use Gaussian process redshift code GPZ \citep{Almosallam2016a, Almosallam2016b}. A final redshift is then obtained from these multiple different redshifts using a hierarchical Bayesian combination framework. The resulting redshifts have very high accuracy, with a median scatter of $\Delta z / (1 + z_{spec})<0.015$ for sources with $z<1.5$. 

\begin{figure}[t]
\centering
    
   \includegraphics[width=\hsize]{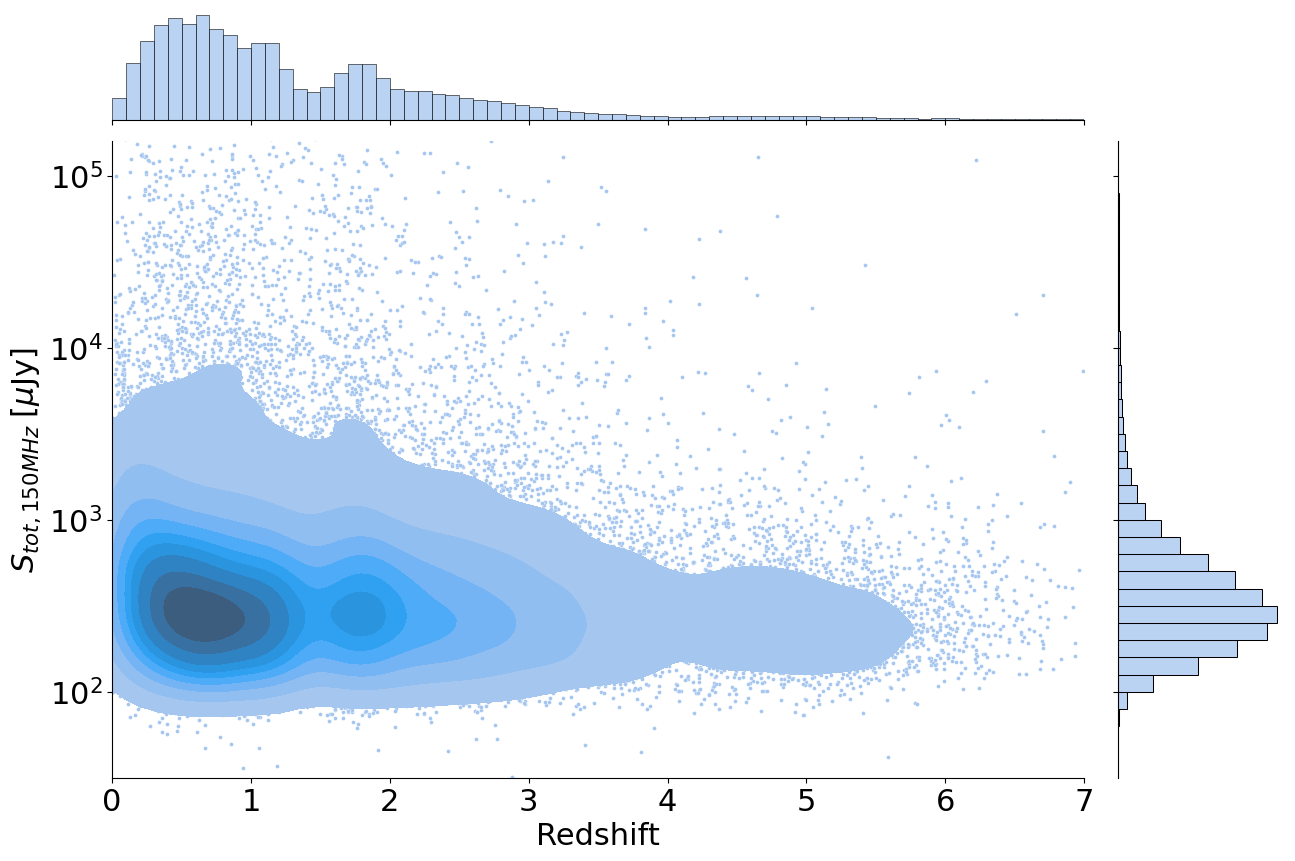}
   \caption{Distribution of the 150 MHz radio flux versus photometric redshift.  Histograms on the side give the distributions of the individual features as well.}
   \label{fig:Data_distribution}
\end{figure}

\begin{figure}[t]
\centering
   \includegraphics[width=\hsize]{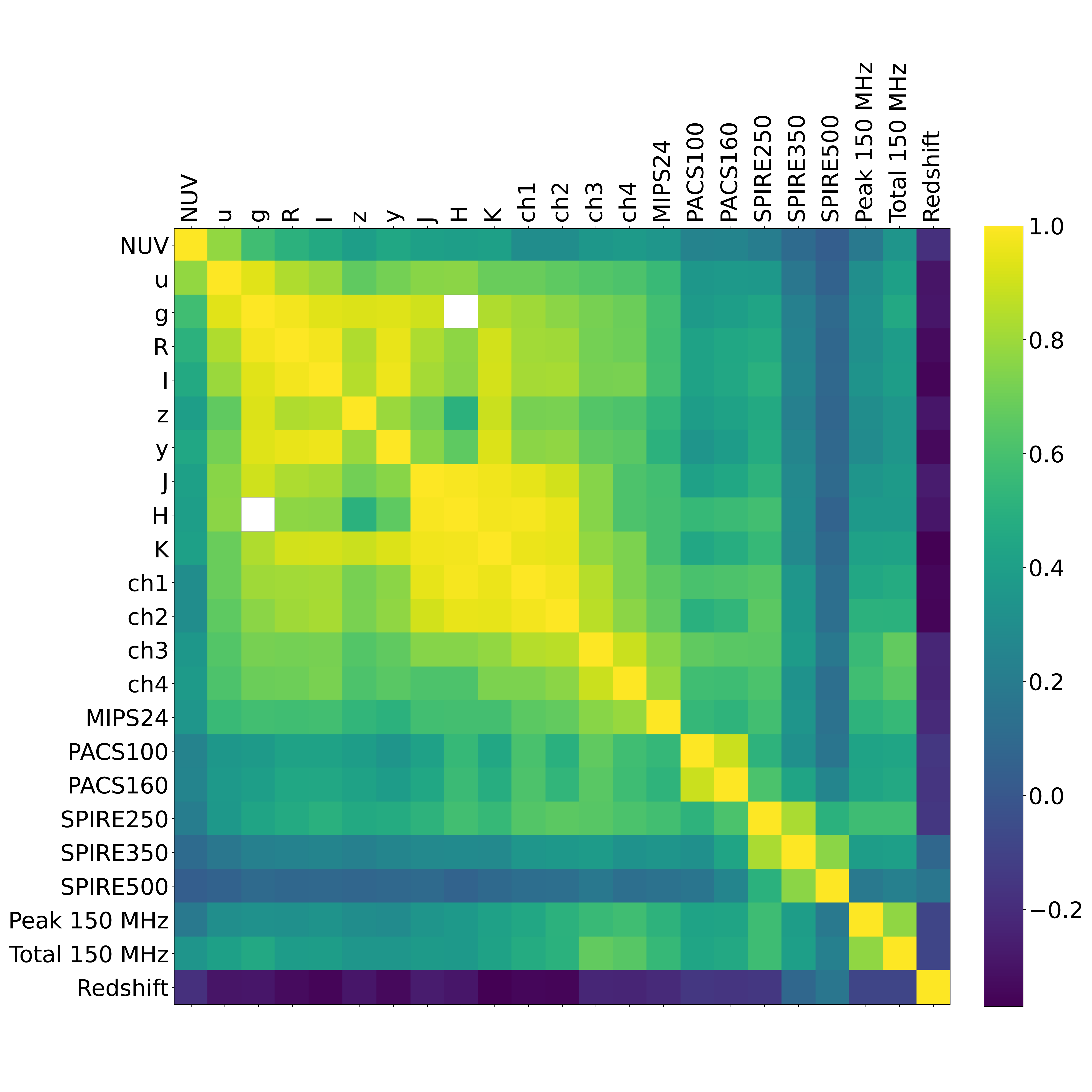}
   \caption{Correlation matrix of all the features used as input for our ML classification. Only SFGs are included in this figure.}
   \label{fig:correlation}
\end{figure}

 To give a general impression of the wide dynamic range of data that is used in this paper, we plot the distribution of the radio 150 MHz flux densities versus redshift in Fig. \ref{fig:Data_distribution}. Data in other wavebands also extend over a large range of redshift and flux densities.
A correlation matrix is plotted for the multi-wavelength photometric data (including the LOFAR radio fluxes) in Fig. \ref{fig:correlation}, which shows how linearly correlated features are. The figure has only been plotted for SFGs since adding AGNs would weaken the correlation between the IR and the radio fluxes. This figure shows, as expected, how fluxes around similar wavelengths (i.e., between the NIR and the MIR) have a stronger correlation.
\subsection{SFG-AGN classification}
\label{sec:AGN_classification}
Using the photometric data and redshifts described in the previous section, \cite{best2023lofar} use four different SED fitting codes to classify sources as different classes of AGN or SFG. We briefly go over each of the SED fitting codes and then discuss the final classification scheme. We refer the reader to \cite{best2023lofar} for details.

The Multi-wavelength Analysis of Galaxy Physical Properties \citep[MAGPHYS;][]{Cunha2008} and Bayesian Analysis of Galaxies for Physical Inference and Parameter EStimation \citep[BAGPIPES;][]{Carnall2018} codes are both SED fitting codes that assume energy balance. This means that the amount of energy absorbed at the optical and UV wavelengths by dust has to be the same as the energy emitted by the dust in the sub-mm and FIR. The main difference in the codes is their implementation of certain parametrisations and models. However, they do generally give consistent results \citep{Pacifici2023}. Unfortunately, neither code includes AGN templates and, thus, cannot provide reliable fits and parameters for galaxies in which the AGN makes a significant contribution to their UV to far-IR flux densities.

Code Investigating GALaxy Emission \citep[CIGALE;][]{Boquien2019} is another model that uses an energy balance approach in SED fitting and modelling. It also has AGN models included, which makes the model significantly better for galaxies with significant AGN emission. The model incorporates the AGN light contribution, the IR emission from the heating of the dust by the AGN, and also the emission in the X-ray. Due to the extra parameters that follow from the AGN-fitting component, the model cannot sample the parameter space of the host galaxy's properties as well as MAGPHYS and BAGPIPES for similar runtimes.

Finally, the version of AGNFITTER \citep{Calistro2016} used by \cite{best2023lofar} does not use the principle of energy balance but instead models four independent emission components. A blue bump, a stellar population, an AGN torus having hot dust emission, and colder dust emission. This way of fitting SEDs works better when the energy balance does not work anymore \citep[e.g., when the UV and FIR emissions are spatially offset from each other;][]{Carnall2018}. It can, however, lead to aphysical solutions or poor constraints on the stellar population parameters.

Using these various models, a set of selection criteria are applied by \cite{best2023lofar} to classify the sources as AGN or SFG and furthermore subdivide them into different AGN classes (HERG, LERG, RQ). To classify a source as a radiative-mode AGN, two of the following three criteria have to be satisfied:

\begin{enumerate}
    \item The 1-sigma lower limit of the AGN fraction (the fraction of IR luminosity from the contribution from the AGN dust torus component; referred to as P16) of the CIGALE fitting is above 0.06 for ELAIS-N1 and Lockman Hole or 0.10 for Bo\"otes.
    \item The P16 value from AGNFITTER is above 0.16 for ELAIS-N1 and Lockman Hole or 0.25 for Bo\"otes.
    \item The lower reduced $\chi^2$ value from MAGPHYS and BAGPIPES SED fits has to be greater than unity and a factor $f$ greater than the lower reduced $\chi^2$ of the CIGALE and AGNFITTER fits. This factor $f$ was 1.36 for ELAIS-N1, 1.59 for the Lockman Hole, and 2.22 for Bo\"otes.
\end{enumerate}

The exact values of these cuts are derived by comparing the classifications to known secure classifications from spectroscopic and X-ray data and from classifications derived from MIR colour-colour diagrams. These criteria mean that a source is classified as a radiative-AGN if the AGN fraction is high in both CIGALE and AGNFITTER or if it only has a high AGN fraction in one of the SED fitting codes but it has a very good SED fit.\\

In addition to classifying sources as radiative-mode or not, \cite{best2023lofar} also classify sources as radio-loud or radio-quiet using the radio data in the LOFAR Deep Fields. These radio-loud AGNs can be identified by analysing the correlation that SFGs have between radio luminosity and their star-formation rate \citep[SFR;][]{Gurkan2018}. Sources that have significantly more radio luminosity than expected from this relation can then be classified as a radio-AGN. \cite{best2023lofar} use a `ridgeline' approach where the sources are binned in narrow redshift bins, and within each bin, the mode of the distribution is picked as a ridgeline point. These ridgeline points can then be fitted with a linear relation. This results in the relation $\log(L_{150\text{MHz}})=22.24+1.08\log(\text{SFR})$, with $L_{150\text{MHz}}$ in W Hz$^{-1}$ and SFR in M$_{\text{sun}}$ yr$^{-1}$. In ELAIS-N1 and the Lockman Hole, a source is deemed an AGN if it exceeds this ridgeline by 0.7 dex (about 3$\sigma$) and by 0.7+0.1$z$ dex for Bo\"otes. The relation is different in Bo\"otes since in that field it is found that the scatter increases at higher redshifts. A small percentage of sources is unclassifiable using this method due to large uncertainties at very low SFRs (below 0.01$M_{\odot}$yr$^{-1}$). Additionally, a few sources are not classified using this method (as they do not reach the radio excess threshold) but are clearly extended ($>$80kpc) multi-component radio sources (incompatible with SFGs) from the LOFAR Galaxy Zoo project \citep{Kondapally2021}. These are added to the sample of radio-loud AGNs (about 0.5\% of the total sample).

Using the two subcriteria of radiative vs non-radiative and radio-loud vs radio-quiet, the four subclasses (SFG, RQ, HERG or LERG) are derived. The results of this class division can be seen in Table \ref{table:class_count}. This table shows that the data has a large imbalance within the classes: the sample contains 20969 AGNs (27\%) and 56640 SFGs (73\%). For supervised ML methods, it can sometimes help to modify the dataset to reduce this imbalance. However, we opt not to do this, as we do not see an improvement in performance. A brief discussion on this can be found in Appendix \ref{app:imbalance}. 
 \begin{table}[t]
\caption{Class count in each field, with information on the number of sources in each of the four classes (SFG, LERG, RQ, and HERG) based on detailed SED analysis using four different SED fitting codes.}
\begin{tabular}{llllll}
\hline
                            & SFG   & LERG  & RQ & HERG     &AGN\\ \hline \hline
ELAIS-N1                    & 23020 & 4342  & 2499  & 387   & 7228    \\
                            & (76\%) & (14\%)  & (8\%)  & (1\%)   & (24\%)    \\
Bo\"otes                    & 12213 & 3219  & 1906  & 391   & 5516  \\
                            & (69\%) & (18\%)  & (11\%)  & (2\%)   & (31\%)    \\
Lockman Hole                & 21407 & 5206  & 2465  & 554   & 8225  \\ 
                            & (72\%) & (18\%)  & (8\%)  & (2\%)   & (28\%)    \\\hline
Total                       & 56640 & 12767 & 6870  & 1332  & 20969 \\ 
                            & (73\%) & (16\%)  & (9\%)  & (2\%)   & (27\%)    \\\hline
\end{tabular}
\label{table:class_count}
\end{table}
\section{Supervised machine learning classification of radio sources}
\label{sec:classification}
Using the data and the labels described in the previous section, we train a supervised ML algorithm on a 2-class scheme (AGN or SFG).
\subsection{Light Gradient Boosting Machine}
For the classification, we use the Light Gradient Boosting Machine \citep[LightGBM or LGBM\footnote{https://github.com/microsoft/LightGBM};][]{lgbm}. LGBM uses a popular machine learning technique called gradient boosting. This ensemble technique uses multiple weaker learners (in LGBM's case, decision trees) to create a better model. Decision trees are structures where at each depth, there is a node that poses a binary decision, for example, if redshift is higher or lower than a given value. This leads to another pair of binary decisions, eventually ending in a classification. This results in 2$^{n-1}$ nodes for a decision tree of depth n. Unlike random forests \citep{breiman_2001}, which split up the dataset with replacement to create multiple decision trees and then combine the results to predict the class, LGBM works sequentially. Each weak learner (a decision tree) is fitted sequentially to reduce the error of the previous model. This loss can then be optimised sequentially using the gradient descent algorithm \citep{Himmelblau1972}; hence the name gradient boosting. The loss we chose to optimise is the log loss, which is defined as 
\begin{equation}
    F = -\frac{1}{N}\sum_{i}^{N}\sum_{j}^{M}y_{ij} \cdot \log(p_{ij}).
\end{equation}
Where $N$ is the number of samples, $M$ the number of different labels, $y_{ij}$ is 1 if the instance belongs to the class and 0 if it does not, and $p_{ij}$ the probability of classifying instance $i$ as label $j$. Gradient boosted decision trees typically result in higher accuracies than random forest \citep{Ping2012}. Contrary to other popular gradient boosting algorithms such as XGBoost \citep{Chen2016}, LGBM grows decision trees per leaf instead of per level (depth-wise). This difference ensures potential higher accuracies but can cause more overfitting.

Another advantage of LGBM over random forests and similar techniques is that it can automatically deal with missing values. Since our data contain missing values for various features, this is an advantageous addition. LGBM uses sparsity-aware split finding, which means that at each decision tree split it assigns a missing value to the side that reduces the loss the most. This allows the algorithm to get, on average, better accuracies on data with many missing values.

Since LGBM is a tree-based method, it is unnecessary to scale or normalise the redshifts and flux densities\footnote{It is possible to add extra features to the model by combining different features into a new feature. Features such as colours could therefore be added. We opt not to do this since the colour information is also contained in the flux densities. The performance of the model does therefore not improve when adding colours.}. However, we have to create training, testing, and validation sets beforehand. This ensures that our model can properly classify the radio sources and does not only learn the structures on the data provided. To create these sets, we use the data we described before. We combine all of our three fields into one large dataset. This dataset is then split 80\%-20\% to create a training and testing set. This testing set is split again 80\%-20\% to create a validation set. The final proportions between training, testing and validation sets are then 80\% (62,087), 16\% (9,934), and 4\% (2,483), respectively. The validation set is used for early stopping of the model. This means that the validation set is evaluated (but not trained on) at each training round of the model. If the performance on this validation set does not improve for ten rounds, the model is stopped. Since the model utilises the early stopping technique, tuning the number of rounds is unnecessary. Normally the number of rounds needs to be tuned to ensure that the model does not train for too long, which reduces performance. However, by using our validation set for early stopping, we can set the number of rounds (n\_estimators in LGBM) to an arbitrarily high number ($10^5$).
\begin{table}[t]
\caption{Hyperparameter search for LGBM. The search parameter space indicates the values in between which the optimal value is sought, and the optimal value is displayed on the right. To find the optimal value, we use a Bayesian optimisation algorithm.}
\label{table:hyperparameters}
\centering
\begin{tabular}{lll}
\hline
                        & Search space     & Optimal value  \\ \hline \hline
num\_leaves             & 10-50            & 46     \\
learning\_rate          & 0.001-0.8        & 0.06369     \\
min\_data\_in\_leaf     & 1-20             & 1     \\
colsample\_bytree       & 0.1-1            & 0.5468        \\
reg\_alpha              & 0-5              & 2.619     \\ 
reg\_lambda             & 0-10             & 7.873     \\ \hline    
\end{tabular}
\flushleft 
\textbf{Notes.} \textit{num\_leaves} is the maximum number of leaves a tree can have. \textit{learning\_rate} determines the step size during the learning process. \textit{min\_data\_in\_leaf} is the minimum amount of samples in each decision leaf. \textit{colsample\_bytree} is the random subset of features the model trains on each iteration. \textit{reg\_alpha} and \textit{reg\_lambda} are L1 and L2 regularisation respectively.
\end{table}

\subsection{Metrics}
\label{sec:metrics}
Proper metrics are necessary to accurately evaluate an ML model. Looking just at the accuracy can give a false impression since it does not take into account the different datasets and performance per class. Therefore we additionally use metrics called the precision, recall, and F$_1$-score per dataset.

The precision, recall, and F$_1$-score are all metrics that range from 0 to 1, with 1 being the best (perfect classification) score, and 0 being the worst. They show the performance of the model per class, instead of overall performance such as accuracy. The precision and recall are defined as, \citep{Olson2008}
\begin{equation}
    \text{Precision}=\frac{\text{TP}}{\text{TP}+\text{FP}},
\end{equation}
and
\begin{equation}
    \text{Recall}=\frac{\text{TP}}{\text{TP}+\text{FN}}.
\end{equation}
Where TP is true positives, FP is false positives and FN is false negatives. For AGNs, TP are the number of AGNs correctly classified as AGNs. FP are the number of classifications where SFGs are wrongly classified as AGNs. FN are the number of classifications where AGNs are classified wrongly as SFGs. For SFGs, the inverse is true for TP, FP, and FN. The precision can then be described as the fraction of sources that are correctly classified as positive, while the recall can be described as the fraction of sources that was recalled. These two metrics can then be combined into the F$_1$-score as: \citep{Olson2008}
\begin{equation}
    F_1=\frac{2}{\text{Recall}^{-1} + \text{Precision}^{-1}}
    =\frac{2\cdot \text{TP}}{2\cdot\text{TP}+\text{FP}+\text{FN}},
\end{equation}
which is the harmonic mean of the precision and the recall. This gives a measure of the accuracy since actual accuracy is not possible class-wise.
\subsection{LGBM hyperparameters}
\label{sec:hyper}
LGBM has a large number of hyperparameters that have to be optimised. These hyperparameters have a large impact on the performance of the model. Instead of trying to find the best parameters `by hand', we use Bayesian optimisation, using the BayesianOptimization python implementation \citep{Fernando2014}. This method tries to optimise a function by generating a posterior distribution. In our case, the function is the performance of the model based on the choice of hyperparameters. As more iterations are run, the posterior distribution improves. The method can then focus on exploring the regions where it expects the output of the function (the model performance) to be the highest. This allows for a much quicker and much more efficient search for the optimal hyperparameters. 

The initial parameter space is taken to be a wide range of values, which can be seen in Table \ref{table:hyperparameters}. We then ran the Bayesian optimisation for 100 iterations. Each iteration cross-validates eight folds and takes the average unweighted F$_1$-score as output. The highest unweighted F$_1$-score is chosen for the optimal hyperparameters, which can also be found in Table \ref{table:hyperparameters}. For a detailed description of these parameters, the LGBM documentation can be consulted. A brief summary of each hyperparameter tuned can be found below Table \ref{table:hyperparameters}. 

\subsection{Overfitting}
\begin{figure}[t]
    \includegraphics[width=8.5cm]{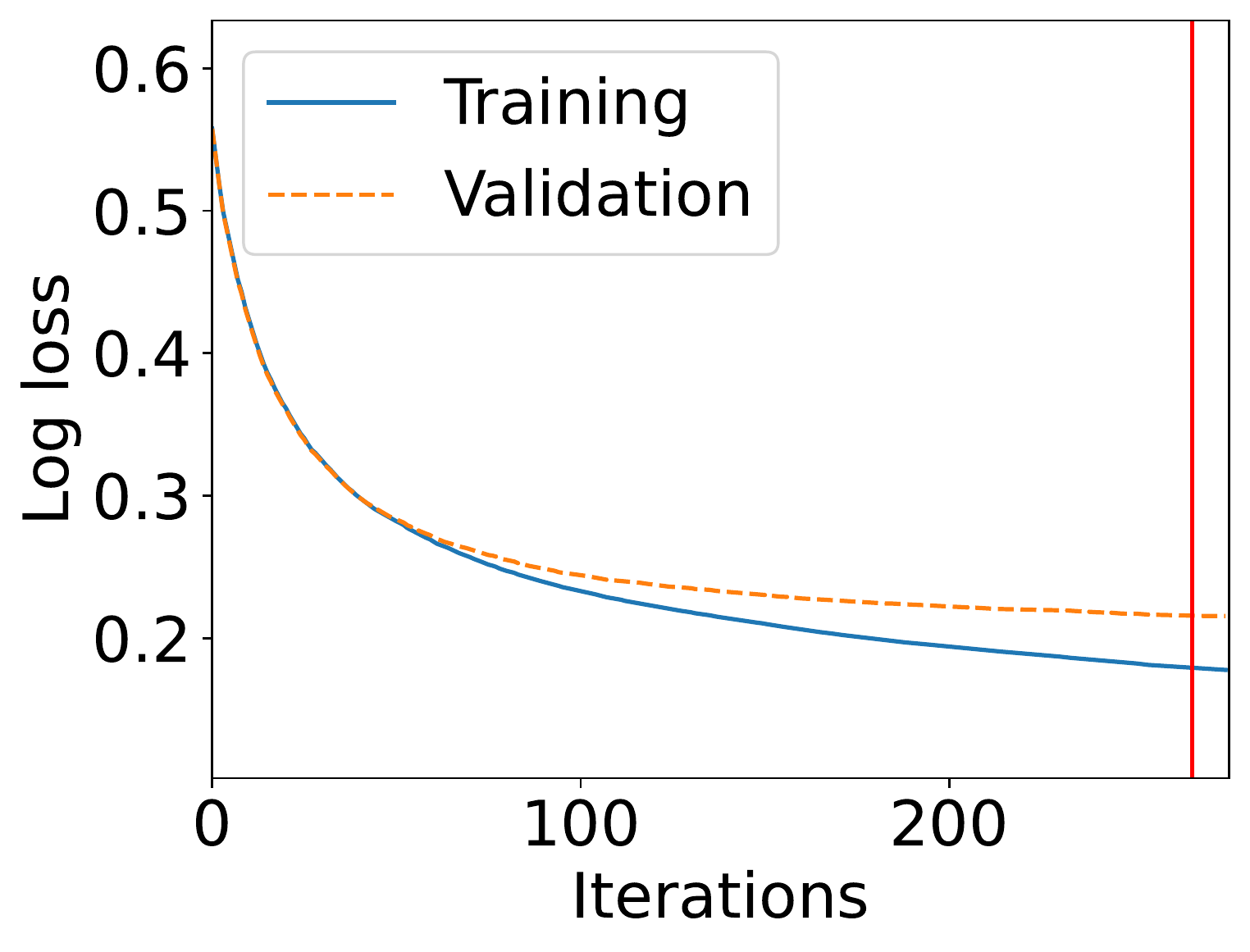}
    \caption{The log loss for the validation and training set for each iteration during training. The training is stopped when the validation loss stopped improving for 10 iterations, indicated by the red vertical line at 266 iterations.}
    \label{fig:history}
\end{figure}
\begin{table*}[t]
\caption{Results of the cross-validated 2-class model. The macro average takes the unweighted mean of the two values above, resulting in a class-balanced metric. The weighted average weighs each value by its fraction in the dataset. The tables display the results of all the data, but also the performance of the test set on the specific fields that we use. Values and errors are derived by the 8-fold cross-validation.}
\centering
\label{table:results}
\begin{tabular}{llll}
\multicolumn{2}{l}{\textbf{All data}}\\
\hline
                 & Precision          & Recall                    & F1-score                \\ \hline
SFG              & 0.92$\pm$0.01  & 0.96$\pm$0.01       & 0.94$\pm$0.01      \\
AGN             & 0.87$\pm$0.02  & 0.79$\pm$0.02        & 0.83$\pm$0.02      \\ \hline
Macro average    & 0.90$\pm$0.02  & 0.87$\pm$0.02       & 0.88$\pm$0.02     \\ 
Weighted average & 0.91$\pm$0.01  & 0.91$\pm$0.01       & 0.91$\pm$0.01      \\ \hline
\end{tabular}
\begin{tabular}{llll}
\multicolumn{2}{l}{\textbf{Lockman Hole}}\\
\hline
                 & Precision          & Recall                    & F1-score                \\ \hline
SFG              & 0.92$\pm$0.01  & 0.96$\pm$0.01       & 0.94$\pm$0.01      \\
AGN             & 0.87$\pm$0.02  & 0.79$\pm$0.03       & 0.83$\pm$0.02      \\ \hline
Macro average    & 0.90$\pm$0.02  & 0.87$\pm$0.03       & 0.88$\pm$0.02     \\ 
Weighted average & 0.91$\pm$0.02  & 0.91$\pm$0.01       & 0.91$\pm$0.01      \\ \hline
\end{tabular}
\vspace{2\baselineskip}\\
\begin{tabular}{llll}
\multicolumn{2}{l}{\textbf{Bo\"otes}}\\
\hline
                 & Precision          & Recall                    & F1-score                \\ \hline
SFG              & 0.91$\pm$0.01  & 0.94$\pm$0.01       & 0.93$\pm$0.01      \\
AGN             & 0.86$\pm$0.02  & 0.80$\pm$0.02       & 0.83$\pm$0.02      \\ \hline
Macro average    & 0.89$\pm$0.02  & 0.87$\pm$0.02       & 0.88$\pm$0.02     \\ 
Weighted average & 0.90$\pm$0.01  & 0.90$\pm$0.01       & 0.90$\pm$0.01      \\ \hline
\end{tabular}
\begin{tabular}{llll}
\multicolumn{4}{l}{\textbf{ELAIS-N1}}\\
\hline
                 & Precision          & Recall                    & F1-score                \\ \hline
SFG              & 0.93$\pm$0.01  & 0.96$\pm$0.01       & 0.94$\pm$0.01      \\
AGN              & 0.87$\pm$0.02  & 0.77$\pm$0.03       & 0.82$\pm$0.03      \\ \hline
Macro average    & 0.90$\pm$0.02  & 0.87$\pm$0.03       & 0.88$\pm$0.02     \\ 
Weighted average & 0.92$\pm$0.01  & 0.92$\pm$0.01       & 0.92$\pm$0.01      \\ \hline
\end{tabular}
\end{table*}
ML models can overfit on the data used for training. Overfitting means that the ML algorithm learns structures on the training data too well and thus performs extremely well on that set, but it is not able to generalise to examples not used during the training. This can mean that the model learns from the noise in the training data, for example. This results in very high performance metrics for the training set but poor performance on the validation set. 

Overfitting can be reduced by tuning the hyperparameters. As mentioned before, we also use a technique called early stopping, where the unweighted average F$_1$-score of the model is evaluated at each training round (epoch) on a set that is not seen during training. If the performance on the validation set does not improve for a certain amount of epochs, the model stops training. Using this method, we can stop the model before it starts overfitting.

To investigate if our model is overfitting we can look at the training histories of the model, since the model runs iteratively we can look at each epoch and see how certain metrics perform on the training and validation set. We use the log loss for this evaluation as this is also the loss the model tries to minimise during training. If there is a large difference between the training and validation set it can indicate overfitting of the model.

Fig. \ref{fig:history} shows the log loss during the training process of the model. When the gap between the training and validation data becomes very wide it is a strong indication of overfitting. The difference between the training and validation data is present but is not that large, remaining within a log loss of 0.1. We can therefore say that the training history does not indicate a significant amount of overfitting of our model.

\section{Results}
\label{sec:results}

\begin{figure*}[t]
    \label{fig:confusion}
    \includegraphics[height=5.1cm]{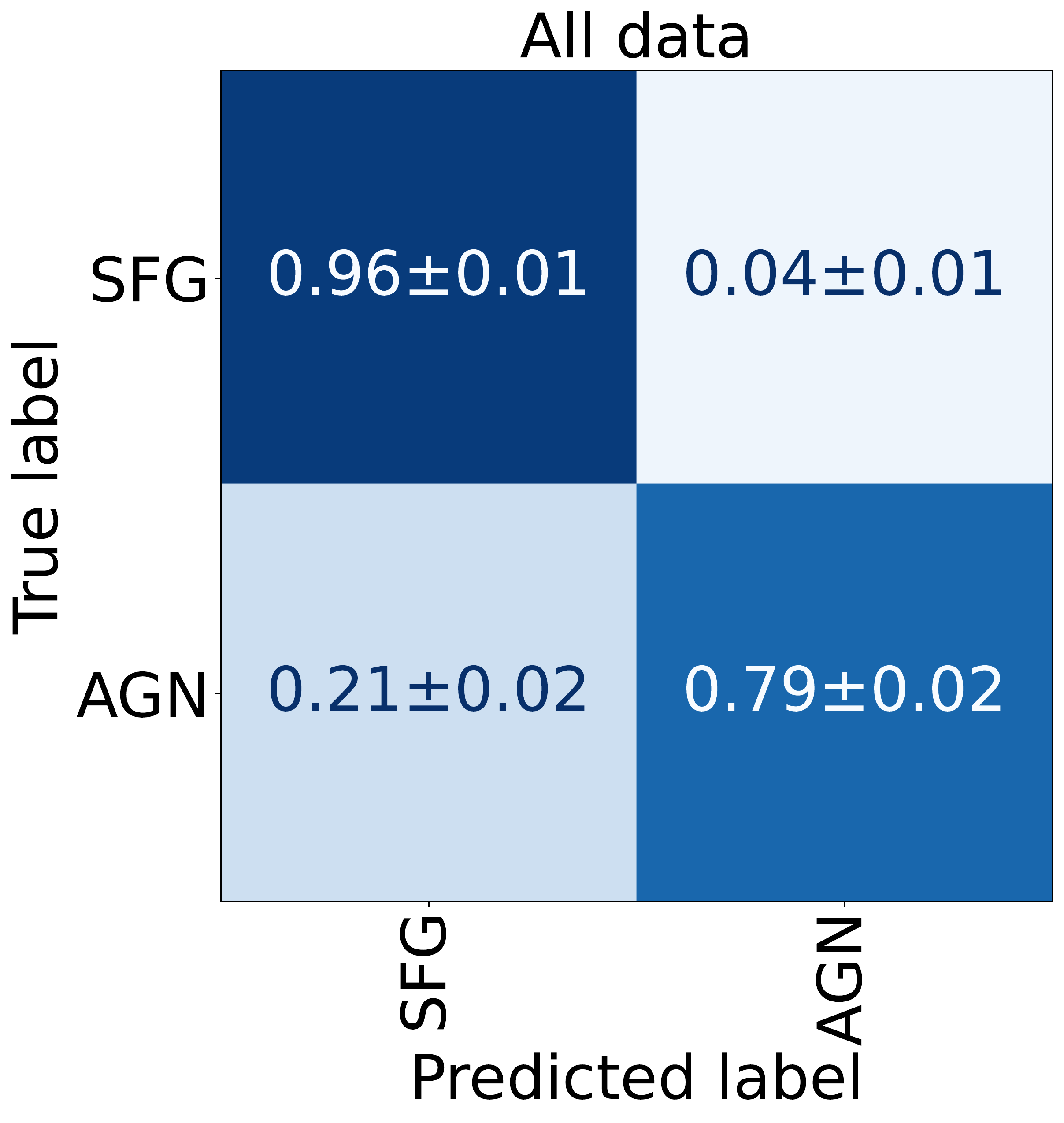}
    \includegraphics[height=5.1cm]{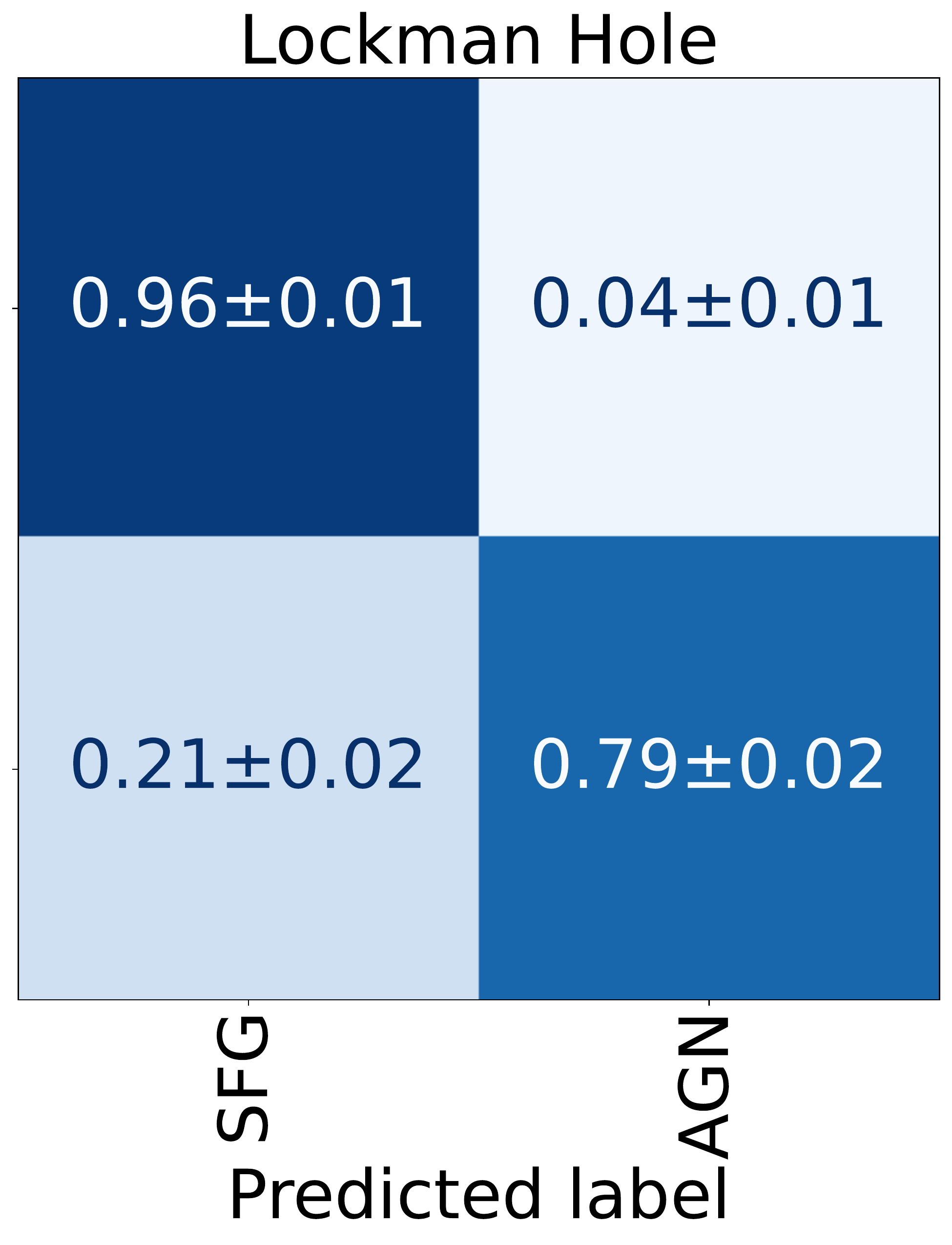}
    \includegraphics[height=5.1cm]{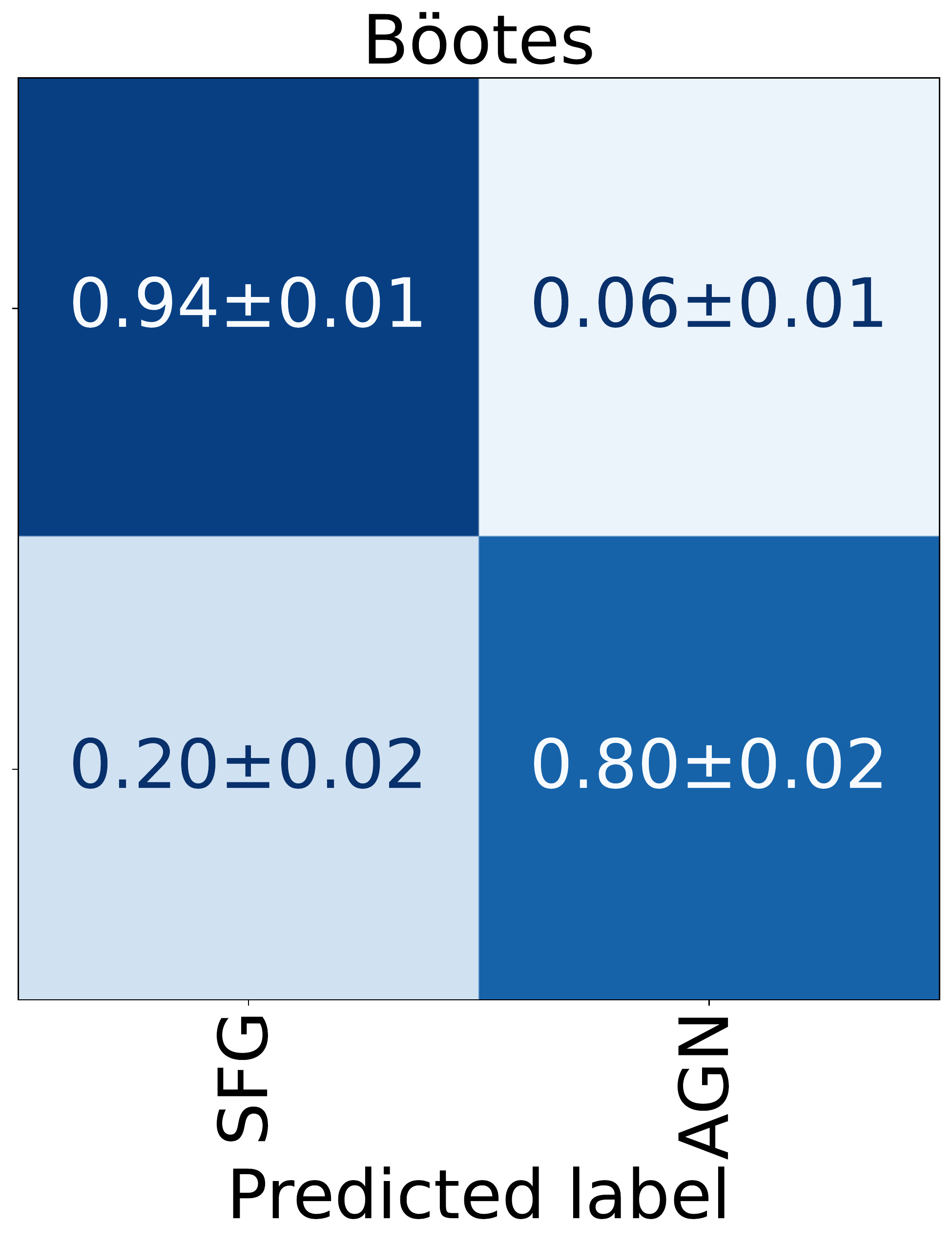}
    \includegraphics[height=5.1cm]{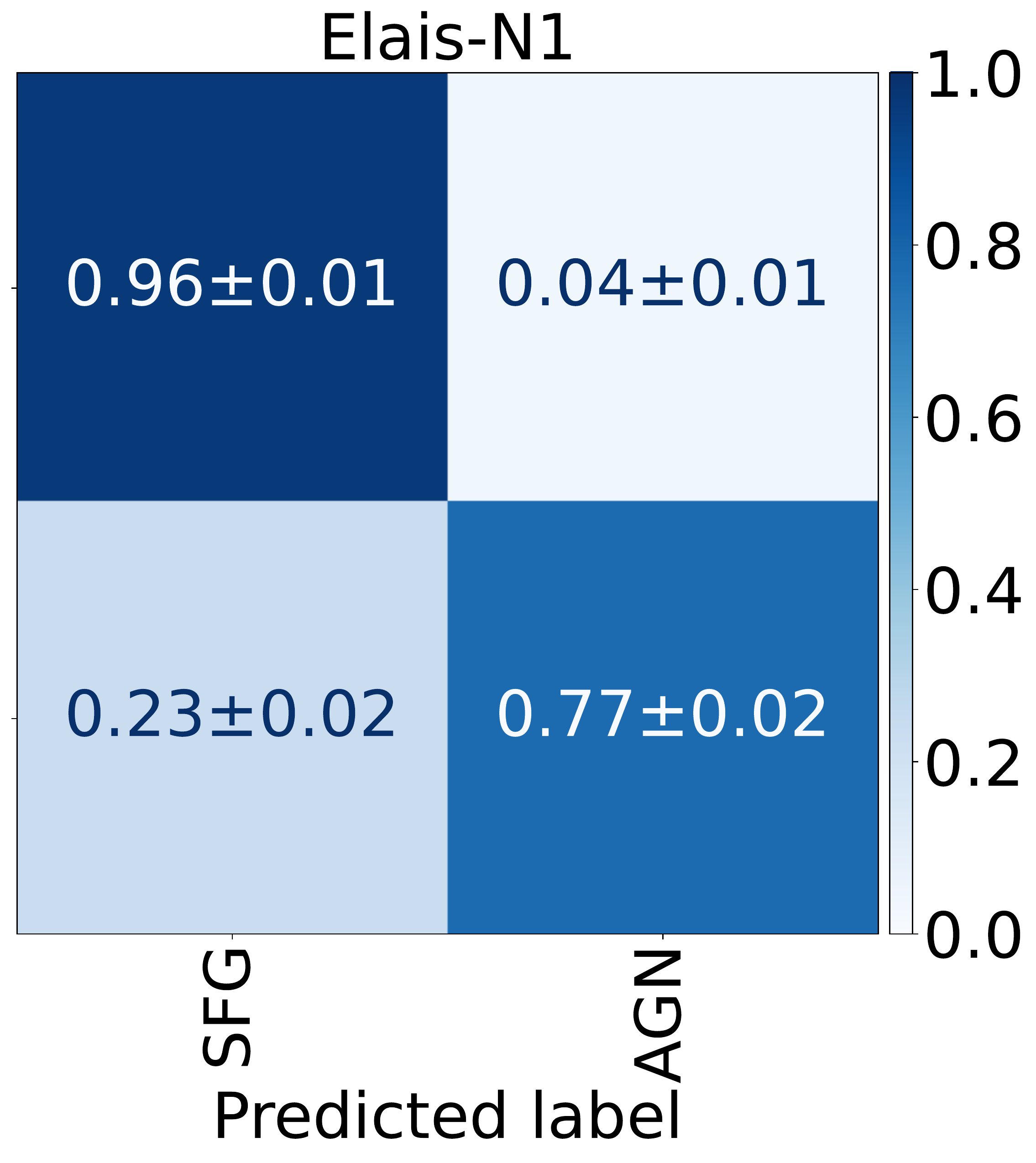}
\caption{Confusion matrices of the cross-validated 2-class model. They show the cross-validated average fraction of how many SFGs and AGNS are classified correctly and wrongly. A perfect classifier has all 1's across the diagonal and 0 everywhere else.}
\end{figure*}
Using the hyperparameters described in Table \ref{table:hyperparameters} we train our ML model. The model is cross-validated in an 8-fold stratified manner, as to keep the distribution between the three fields the same. In this section, we analyse how our trained model performs, for that averages and 1-$\sigma$ standard deviations are calculated and analysed.
\subsection{Overall performance}
Our model is trained on a binary classification scheme (AGN vs SFG). \cite{best2023lofar} however, provide four classes (HERG, LERG, RQ, or SFG) for their source classification, which means the model can also be trained on four classes instead of two. We decide to focus on two classes in this paper since the performance on the 4-class model is poorer. Even though the 4-class model shows similar accuracy as the 2-class model described later in this section, the performance on the minority classes is very poor. Particularly for the HERGs, the classifier reached very low ($<50$\%) precision and recall. This bad performance is mostly due to the low number of sources in some classes. In Appendix \ref{app:4class}, we summarise our investigations of a 4-class model. In the main paper, we focus on the 2-class model.

For the 2-class model Table \ref{table:results} shows that our classifier has a total accuracy of 91\%, which is a very good performance. This value is only representative of our class distribution (AGN or SFG); this value can be heavily biased if there is a large class imbalance. In our case, our sample contains a large number of SFGs, which means that they influence the accuracy more than the AGNs. The larger fraction of SFGs influences the loss function of the model while training and thus results in a better performance for them compared to the AGNs. For the AGNs, a precision of 87\% is measured, which means that 87\% of the sources classified as AGNs are true AGNs according to the labels used in this work. The recall is lower at 79\%, meaning that we recover 79\% of the AGNs in the data.  The overall performance of the model can better be evaluated by looking at the unweighted (macro) averages of the metrics. The unweighted average is a good metric since it is not impacted by class imbalance. The macro F$_1$-score, which is a combination of both precision and recall is 88\%. Confusion matrices are plotted in Fig. \ref{fig:confusion}. These show the fractions of how the model classifies sources. These have been normalised over the rows, such that the diagonal represent the recalls. An ideal classifier has all 1's across the diagonal and 0's on the off-diagonal squares. We can see that even for the minority class (the AGN), the performance holds up quite well, although about 22\% of the AGNs are misclassified as SFGs. 

The performance of the three classes of AGNs can also be analysed. This analysis is still done on the two-class model, we simply look at how well the subclasses are classified as AGN or not. The recall is 93\%$\pm$3\% for HERGS, 70\%$\pm$3\% for RQ and 81\%$\pm$2\% for LERGS. Compared to the class distribution (12767, 6870 and 1332 for LERGs, RQs and HERGs respectively) the HERGs seem to overperform as would be expected from their class size. 

The analysis above is about the performance of our model using the three fields as training, validation and testing sets. We are, however, also interested in the performance of the model on a new, unseen dataset. We can simulate such a new dataset, by only using two fields as training validation and testing sets, while using another field purely as a testing set. We can then compare the performances between the two testing sets to see if the model can generalise to new data. We use the Lockman Hole and Bo\"otes data for our 2 fields and ELAIS-N1 as our testing field. The performance on the testing set when training on the 2 fields was approximately 2\% lower in precision and recall and approximately 1\% lower in accuracy. This indicates that our model would perform well on new data, provided the quality was similar to ELAIS-N1.

\begin{figure}[t]
    \includegraphics[width=8.5cm]{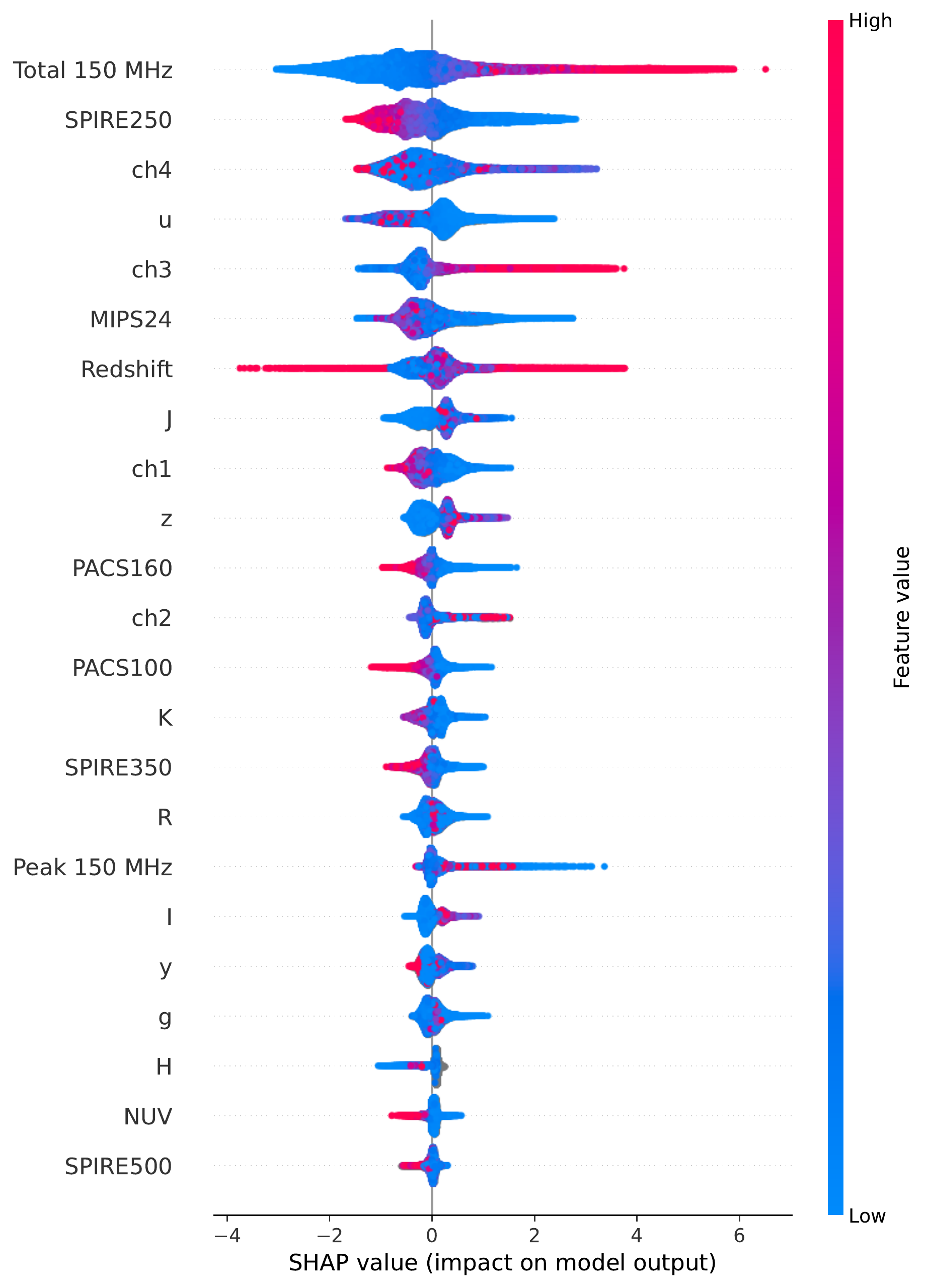}
    \caption{Feature importance using SHAP values. The features are ordered by importance from top to bottom, with the most important feature being on top. On the x-axis the SHAP value is displayed, a positive value indicates a higher chance of the associated source being an AGN, while a negative value is a higher chance of the source being an SFG. The value of the feature is shown via the colour, which is also displayed on the right in a colour bar. For instance a higher radio flux results in a higher probability of the source being an AGN.}
    \label{fig:feature_relevance}
\end{figure}
In addition to analysing the performance metrics, the importance of the individual features can also be investigated. This not only helps identify which features are more important but also gives a better view of how all the features impact the model overall. Various methods exist for looking at feature relevance, usually relying on some kind of score that each feature gives. We use SHapley Additive exPlanations \footnote{https://github.com/slundberg/shap} (SHAP) values \citep{Lundberg2017}. SHAP gives each feature a value that describes its importance in the model. Additionally, it can show how features impact the model by looking at the size and sign of the value. A bigger and positive value means a higher impact and a lower and negative value means a lower impact on the classification. Using the Python package created by \cite{Lundberg2017}, the SHAP values for the model are found. In Fig. \ref{fig:feature_relevance}, we show the feature importance, where a higher SHAP value means it is more likely to be an AGN. The figure shows mostly expected results. Radio and IR features are generally the most important, while fluxes in the visible are less important. Furthermore, higher radio fluxes result in the source being more likely to be an AGN, which is expected for the radio-loud AGNs. This figure does not show any cross-interactions between the different features, it only displays how overall one feature impacts the classifier. 
\subsection{Dependence on sample size, SED sampling, and signal-to-noise}
\label{sec:redshift}
The training data have more samples at lower redshift than at higher redshift. The model, therefore, learns the underlying structures at lower redshifts better since machine learning algorithms perform better with more data. Additionally, lower redshift sources generally have higher-quality data than higher redshift sources. This means that in addition to the sample-size dependence mentioned above, the metrics are worse due to reduced data quality. SED classifications also become less reliable above z=2.5 \citep{best2023lofar}. These three effects combined indicate that the performance degrades relatively quickly with increased redshift.
\begin{figure}[t]
    \includegraphics[width=8.5cm]{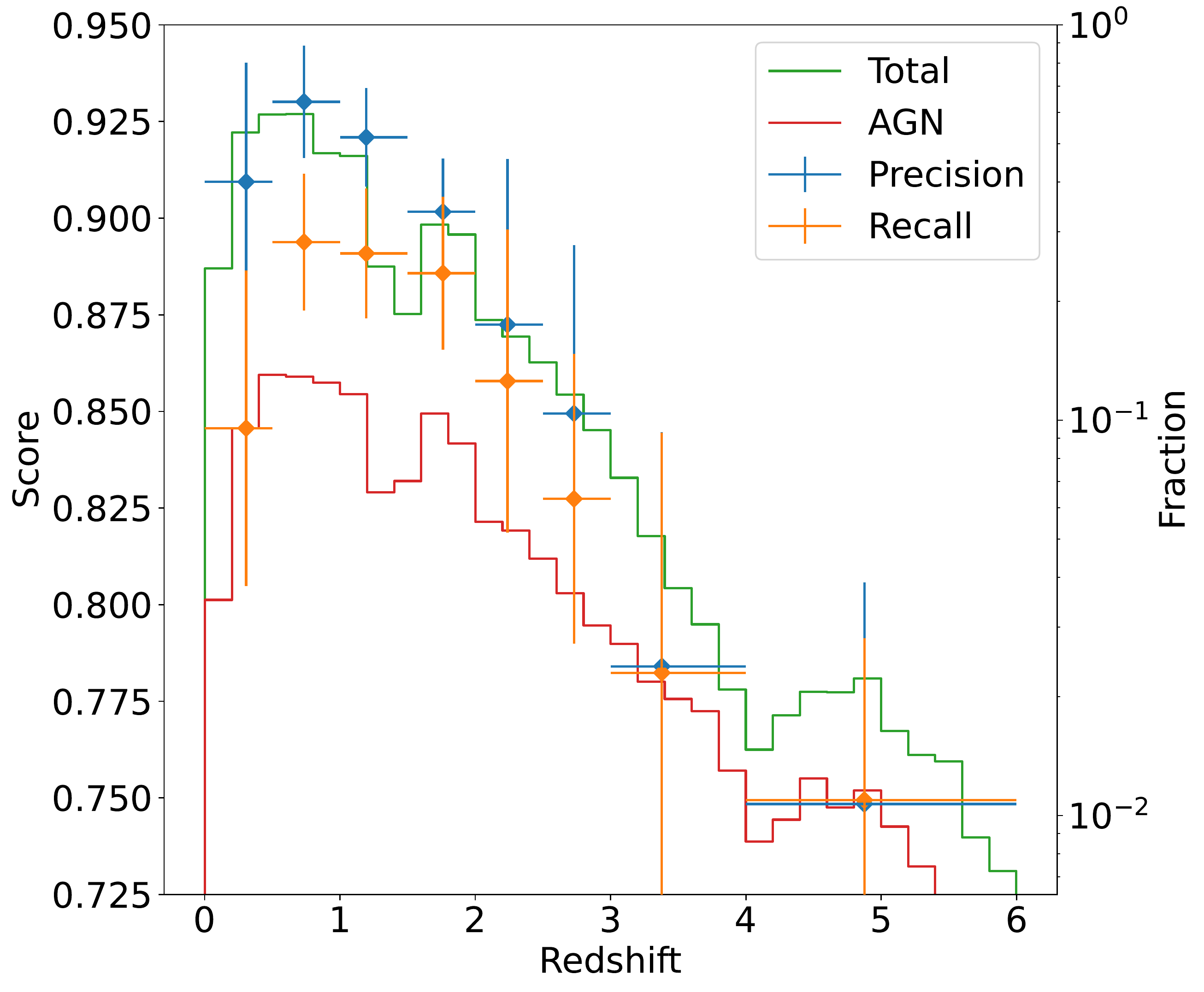}
    \caption{8-fold cross-validated results of binned testing sets based on redshift. On the x-axis, the redshift is displayed. The points and errors are calculated by taking each bin's mean and boundaries. In addition to the precision and recall on the left y-axis, the fraction of the data contained within the bin is plotted on the right y-axis. The y-errors represent 1-$\sigma$ standard deviations of the scores. The borders of the bins are [0, 0.5, 1, 1.5, 2, 2.5, 3, 4, 6].}
    \label{fig:redshift}
\end{figure}
To measure these effects, we perform 8-fold cross-validated tests in which we bin testing data by redshift and then measure the performance on them. These results are plotted in Fig. \ref{fig:redshift}, where we also plot the corresponding fraction in a histogram. This plot clearly shows that the training size and score decrease when the redshift increases. The bin sizes have been chosen manually such that each bin contains at least 5\% of the training data. 

Our model has been trained on data with a certain amount of features (fluxes and redshift). Other datasets, however, may not have the same features as the ones our model has been trained on. It is, therefore, important to analyse how well our model performs when a testing set has fewer features.

LGBM does automatic missing value imputation. This technique is convenient when there are some missing values, but it does not perform well when an entire feature (i.e., an entire column in the data) is missing. This is investigated in some detail in Appendix \ref{app:missing}. Therefore, if we want to evaluate how well a model performs if a feature is missing, we cannot simply drop a column from the testing set and then evaluate the metric. Instead, we have to retrain the model with this column missing and evaluate the performance. We cannot give all possible combinations of missing features since if we have n features, the number of all possible combinations is $2^n$ (the powerset) \citep{Halmos1960}, for n features, which is extremely large in our case. Instead, we focus on some relatively common combinations and some combinations which have a lot of impact on the performance. 
\begin{figure}[t]
    \includegraphics[width=8.5cm]{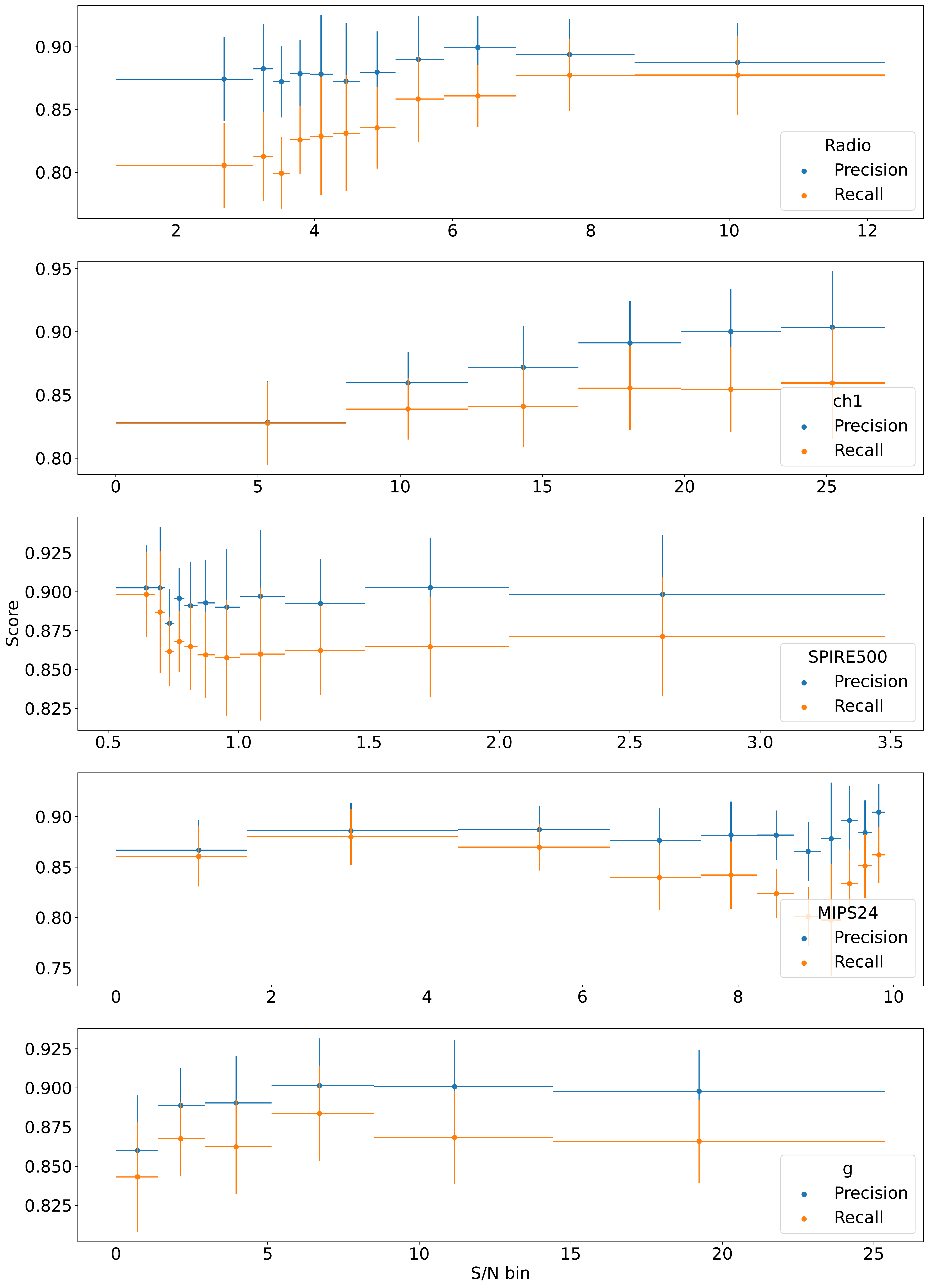}
    \caption{Performance per S/N bin. The x-axis shows the mean value of the S/N bin. The y-axis denotes the macro average precision and recall. The uncertainties in the y-axis are calculated from the 1-$\sigma$ standard deviation over the 8-fold cross-validation. The uncertainty in the x-axis is the bin width.}
    \label{fig:SN}
\end{figure}

Once again, we use 8-fold cross-validation to calculate metrics. We remove the features from the training, validation, and testing set for each missing feature selection and then measure the performance. The results of this can be seen in Table \ref{table:less_features}. It can be seen that the model's performance does not degrade too much unless we remove a large number of features. This means that using this model with very few features is not recommended due to its bad performance. 
\begin{table*}[t]
\caption{Model performance when retraining the model on fewer features. The model is trained on an 8-fold cross-validation, and the precision, recall for each class and the total macro F$_1$-score are calculated. In this table MIPS refers to MIPS24, PACS to PACS 100 and PACS160 and SPIRE to SPIRE250, SPIRE350 and SPIRE500.}
\label{table:less_features}
\centering
\begin{tabular}{llllll}
\hline
                                                & Precision SFG & Recall SFG    & Precision AGN & Recall AGN & F$_1$-score       \\\hline \hline
All                                             & $0.92\pm0.01$ & $0.96\pm0.01$ & $0.87\pm0.02$ & $0.78\pm0.02$ & $0.88\pm0.02$     \\\hline

NUV, U, \textit{grizy}, J, H, K, ch1-ch4, MIPS, \\PACS, SPIRE & $0.95\pm0.01$ & $0.89\pm0.01$ & $0.69\pm0.01$ & $0.83\pm0.02$ & $0.83\pm0.01$     \\\hline

NUV, U, J, H, K, ch1-ch4, MIPS, \\PACS, SPIRE                    & $0.94\pm0.01$ & $0.89\pm0.01$ & $0.68\pm0.01$ & $0.82\pm0.02$ & $0.83\pm0.01$     \\\hline

NUV, U, \textit{grizy}, J, H, K, ch3, ch4, \\MIPS, PACS, SPIRE                         & $0.95\pm0.01$ & $0.89\pm0.01$ & $0.68\pm0.01$ & $0.82\pm0.01$ & $0.83\pm0.01$     \\\hline

NUV, U, \textit{grizy}, J, H, K, MIPS, PACS, \\SPIRE               & $0.94\pm0.01$ & $0.87\pm0.01$ & $0.62\pm0.01$ & $0.81\pm0.02$ & $0.80\pm0.02$     \\\hline

NUV, U, \textit{grizy}, J, H, K, PACS, SPIRE     & $0.94\pm0.01$ & $0.85\pm0.01$ & $0.56\pm0.01$ & $0.77\pm0.01$ & $0.77\pm0.01$     \\\hline

NUV, U, \textit{grizy}, J, H, K, MIPS, SPIRE                              & $0.95\pm0.01$ & $0.87\pm0.01$ & $0.61\pm0.01$ & $0.80\pm0.02$ & $0.80\pm0.01$     \\\hline

NUV, U, \textit{grizy}, J, H, K, MIPS, PACS                  & $0.94\pm0.01$ & $0.85\pm0.01$ & $0.55\pm0.01$ & $0.77\pm0.01$ & $0.77\pm0.01$     \\\hline

NUV, U, \textit{grizy}, J, H, K                  & $0.94\pm0.01$ & $0.81\pm0.01$ & $0.40\pm0.01$ & $0.72\pm0.02$ & $0.69\pm0.02$     \\\hline
\textit{grizy}                  & $0.94\pm0.01$ & $0.79\pm0.01$ & $0.34\pm0.02$ & $0.71\pm0.02$ & $0.66\pm0.02$     \\\hline

NUV, \textit{grizy}, J, H, K, ch1-ch4, MIPS, \\PACS, SPIRE, 150 MHz                                               & $0.96\pm0.01$ & $0.92\pm0.01$ & $0.78\pm0.01$ & $0.87\pm0.02$ & $0.88\pm0.01$     \\\hline

NUV, U, \textit{grizy}, ch1-ch4, MIPS, PACS, \\SPIRE, 150 MHz                                         & $0.96\pm0.01$ & $0.92\pm0.01$ & $0.78\pm0.01$ & $0.87\pm0.02$ & $0.88\pm0.01$     \\\hline

NUV, U, J, H, K, ch1-ch4, MIPS, PACS, \\SPIRE, 150 MHz                                   & $0.96\pm0.01$ & $0.92\pm0.01$ & $0.77\pm0.01$ & $0.86\pm0.01$ & $0.88\pm0.01$     \\\hline

NUV, U, \textit{grizy}, J, H, K, MIPS, PACS, \\SPIRE, 150 MHz               & $0.95\pm0.01$ & $0.90\pm0.01$ & $0.71\pm0.01$ & $0.85\pm0.02$ & $0.85\pm0.01$     \\\hline

NUV, U, \textit{grizy}, J, H, K, ch3, ch4, MIPS, \\PACS, SPIRE, 150 MHz                                        & $0.96\pm0.01$ & $0.92\pm0.01$ & $0.78\pm0.01$ & $0.86\pm0.02$ & $0.88\pm0.01$     \\\hline

NUV, U, \textit{grizy}, J, H, K, ch1, ch2, MIPS, \\PACS, SPIRE, 150 MHz                                        & $0.96\pm0.01$ & $0.91\pm0.01$ & $0.74\pm0.01$ & $0.86\pm0.02$ & $0.86\pm0.02$     \\\hline
    
NUV, U, \textit{grizy}, J, H, K, MIPS, PACS, \\SPIRE, 150 MHz                              & $0.95\pm0.01$ & $0.90\pm0.01$ & $0.72\pm0.01$ & $0.85\pm0.02$ & $0.86\pm0.02$     \\\hline

NUV, U, \textit{grizy}, J, H, K, 150 MHz                           & $0.94\pm0.01$ & $0.84\pm0.01$ & $0.53\pm0.01$ & $0.77\pm0.02$ & $0.76\pm0.02$     \\\hline

NUV, U, \textit{grizy}, J, H, K, ch1-ch4, PACS, \\SPIRE, 150 MHz                                          & $0.95\pm0.01$ & $0.91\pm0.01$ & $0.76\pm0.01$ & $0.85\pm0.01$ & $0.87\pm0.02$     \\\hline

NUV, U, \textit{grizy}, J, H, K, ch1-ch4, MIPS, \\SPIRE, 150 MHz                                & $0.96\pm0.01$ & $0.92\pm0.01$ & $0.78\pm0.01$ & $0.87\pm0.02$ & $0.88\pm0.02$     \\\hline

NUV, U, \textit{grizy}, J, H, K, ch1-ch4, MIPS, \\PACS, 150 MHz                    & $0.95\pm0.01$ & $0.91\pm0.01$ & $0.74\pm0.01$ & $0.84\pm0.02$ & $0.85\pm0.01$     \\\hline

NUV, U, \textit{grizy}, J, H, K, ch1-ch4, 150 MHz                            & $0.94\pm0.01$ & $0.88\pm0.01$ & $0.67\pm0.01$ & $0.82\pm0.02$ & $0.82\pm0.02$
\\\hline    
\end{tabular}
\end{table*}

Lastly, the quality of the data can have a significant impact on the performance of the classifier. The quality of the data is  measured by signal-to-noise ratios (S/N). We calculate the S/N by dividing the fluxes by the errors provided with the multi-wavelength data. To measure the model's performance for different S/N, we take binned S/N cuts in a particular waveband in the testing set and see how performance differs for each bin. The model is trained on all the bins simultaneously to compare the different bins' performance fairly.

To ensure that we can compare models fairly and objectively, we have to ensure that the main difference between each bin is the S/N and is not dependent on other factors. We, therefore, use adaptive bin sizes. This is done to ensure that each bin has the same amount of sources in the training set. In general, this results in the lower S/N bins being relatively narrow and the higher S/N bins being wider since the sample size peak happens at a relatively lower S/N. Each bin has a sample size of 5000. Since our total sample size is not a multiple of 5000, we discarded some very high S/N sources.
\begin{figure}[t]
    \includegraphics[width=8.5cm]{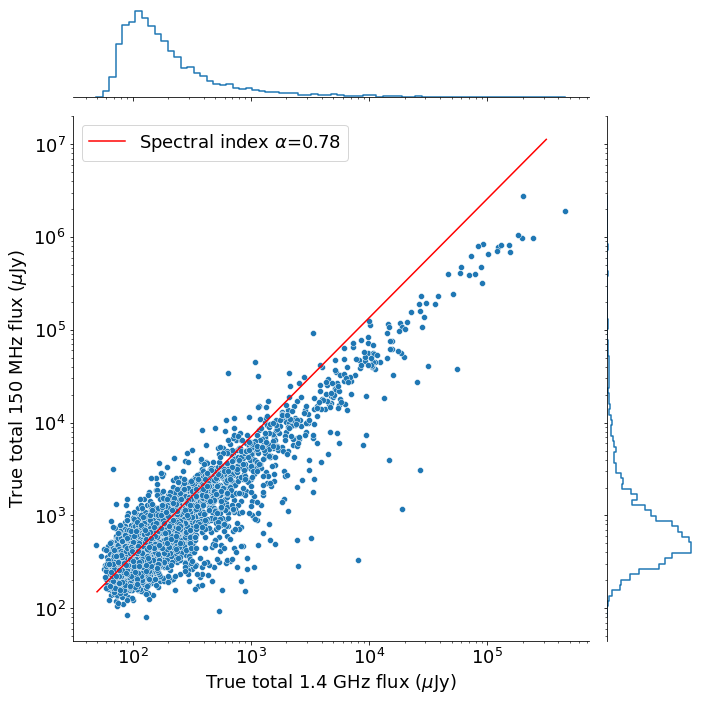}
    \caption{Flux comparison of the observed LOFAR 150 MHz data and the observed 1.4 GHz data. A spectral index of 0.78 derived by \cite{Mahony_2016} is plotted in the red line.}
    \label{fig:radio_conversion}
\end{figure}

Because some of the flux densities are highly correlated, an increase of S/N in some bands increases the S/N of many bands simultaneously, as the noise is largely uncorrelated. This means that the differences between S/N bins are significantly larger for these flux densities, while the performance difference might be minimal for other flux densities. The correlation between the different S/N of the features can be inferred from the correlation matrix in Fig. \ref{fig:correlation}.

Using the abovementioned precautions, we ran the model on an 8-fold cross-validation and measured some of the features' macro-average precision and recall scores. We chose a selection of features that showed limited linear correlation to investigate most of the spectrum. The results have been plotted in Fig. \ref{fig:SN}. This figure shows that for certain bands such as the radio and IRAC channel 1, an increase in S/N results in a better performance of the model. For the g band, a positive trend is less significant but still visible. For the IR features, an increase in S/N does not indicate an increase in performance, with even a possible decrease. This is contrary to expectations but could be due to uncertainties in the error estimates of these features.

\subsection{Application to radio galaxies detected at 1.4 GHz}
\begin{table}[t]
\caption{Model performance when trained with converted 1.4 GHz data.}
\centering
\label{table:results_14}
\begin{tabular}{llll}
\hline
                    & Precision & Recall    & F1-score  \\ \hline
SFG                 & 0.89      & 0.90      & 0.89      \\
AGN                 & 0.86      & 0.85      & 0.85      \\ \hline
Macro average       & 0.87      & 0.87      & 0.87     \\ 
Weighted average    & 0.88      & 0.88      & 0.88      \\ \hline
\end{tabular}
\end{table}
Since much research in radio astronomy is done at 1.4 GHz, we also include a brief analysis of the performance of using the 1.4 GHz radio data instead of the 150 MHz radio data. We use the 1.4 GHz data available in the Lockman Hole from the Lockman Hole Project \citep{Prandoni2018}. This set is cross-matched using a 3" matching radius to the LOFAR Lockman Hole  sample, resulting in 4005 matches. We compare the performance of the model using the LOFAR 150 MHz data vs predicted 150 MHz radio data from 1.4 GHz Lockman Hole project radio data. The predicted 150 MHz radio data are generated from the 1.4 GHz radio using a simple spectral index of $\alpha=0.78$ derived by \cite{Mahony_2016}. They note that this index becomes steeper with increasing flux densities (from $\alpha=0.75$ to $\alpha=0.84$). We do not change our spectral index with flux density since detailed spectral index analysis is often not available on real data. The effect of this simpler approach can be seen in Fig. \ref{fig:radio_conversion}, where the spectral index fits poorer at higher flux densities.

To ensure that our model does not train on the predicted 150 MHz fluxes, we simply use this new sample of 4005 sources as a testing set while training on the rest of the data. We compare the performance of this set with the true 150 MHz fluxes. The performance of the original 150 MHz sample reaches an accuracy of 91\% and a macro average F$_1$-score of 90\%, similar to the expected values we know from Table \ref{table:results}. The performance of the model with the converted 1.4 GHz radio fluxes can be seen in Table \ref{table:results_14}. The accuracy is 88\%, and the macro average F$_1$-score is 87\%. Therefore, the performance is slightly worse, but it is still much better than simply dropping the radio features altogether (which results in a macro F$_1$-score of 83\%). 
\section{Conclusions}
\label{sec:conclusions}
To conclude, in this paper, we create a supervised ML model to classify sources detected in extragalactic radio surveys as AGN or SFG. We use extensive radio and multi-wavelength data in three LoTSS Deep Fields: ELAIS-N1, Lockman Hole and Bo\"otes. Each field also has high-quality photometric redshifts. We combine these three fields by selecting features that are available in most of the fields resulting in 77,609 sources, of which 20,969 are AGNs. 

We create the ML classifier by using a decision-tree-based algorithm called LGBM. The 8-fold cross-validated testing resulted in an $F_1$-score of 0.94$\pm$0.01 for SFGs and 0.83$\pm$0.02 for AGNs, resulting in an average macro F$_1$-score of 0.88$\pm$0.06 and an accuracy of 0.91$\pm$0.01. We did not find significant deviations in the performance in each of the three different fields. The biggest source of error in the model is that a fraction of 0.22$\pm$0.02 of the AGNs is misclassified as SFGs. 

We test the model's performance for different sample sizes at different redshift bins. We find that lower sample sizes in the training set in redshift bins result in reduced performance, which is expected. Furthermore, we investigate the model's performance when fewer features are used. We retrain models with features removed and then test the performance. We find that performance decreases more when IR and radio features are removed while removing the visible and UV features barely reduces performance. Lastly, using S/N bins to see the model's performance for different S/N values, we find that, in general, higher S/N results in some improvement of the model performance.

A public release of the model is available at \url{https://github.com/Jesper-Karsten/MBASC}. This allows other researchers to use the model to classify AGNs in their own datasets.

\bibliography{aanda.bib}
\appendix{}
\section{Data imbalance}
\label{app:imbalance}
As can be seen in Table \ref{table:class_count}, the class sizes differ. The SFGs account for about 66\% of the complete dataset. Imbalanced datasets can impact the performance of the model. This is because the model then learns more from the majority class while not learning much from the minority classes. Additionally, it makes analysis of the model harder since most performance metrics are mostly influenced by the majority class.

A simple option to remedy this is to assign class weights to sources based on their class, such that a class that is x times more frequent than some other class has a weight of 1/x. This weight then impacts the score that the algorithm calculates for its gradient descent. LGBM provides a simple sample\_weight argument for this. Unfortunately, when using this parameter in the Bayesian optimisation, it reduces both accuracy and F$_1$-score by about 1\%. This is not unusual since sometimes adding an extra sample weight can make the model less able to learn on the larger classes.

Another relatively simple approach is to remove sources such that the data are more balanced. This option is not viable in our case because removing sources harms the model's performance more (due to having fewer data to train on) than it improves due to a more balanced class distribution.

Lastly, we also try generating extra data. We do this by using a relatively simple approach where we generate additional sources using the data we already have. We generate new sources by using Normal distributions with the errors on each flux and the redshift as the standard deviation. Unfortunately, when carefully generating new data such that all classes are balanced, the model's performance does not improve. The lack of improvement can be explained by the fact that these Gaussian-generated sources are still quite similar to the original sources. They do not convey new complex information about what these sources could be. This approach might actually increase the amount of overfitting we do on the data since it essentially adds noisy copies of the data to the training set.

\section{Missing values}
\label{app:missing}
Even though LGBM allows the automatic handling of missing values, this feature does not work well with the entire missing features (columns) in the data. This fact can be shown qualitatively by manually removing values in certain columns, measuring the performance on the same model, and comparing this against a retrained model where the same columns have been dropped in the training and validation set. The results are shown in Table \ref{table:remove_columns}. These results are not cross-validated since retraining the model each time is very time-consuming. Still, they do show qualitatively that there is a lot of improvement for most features when retraining the model in general. For this reason, we employ a strategy of retraining the model when testing performance for missing features. This approach is, however, not always necessary, as for some features the improvement is minimal.
\begin{table}
\caption{Comparison of macro F$_1$-score performance of the model by different missing features strategies. The left column shows a model which has not been retrained after removing one or multiple features, while the right column has been retrained.}
\label{table:remove_columns}
\centering
\begin{tabular}{lll}
\hline
                                & Only dropped  & Retrained  \\ \hline \hline
No missing                      & 0.88          & -          \\
LOFAR 144 MHz                   & 0.77          & 0.84       \\
u                               & 0.86          & 0.88       \\
J, H, K                         & 0.87          & 0.88       \\
g, r, i, z, y                   & 0.86          & 0.88       \\ 
ch1, ch2                        & 0.83          & 0.88       \\
ch3, ch4                        & 0.83          & 0.87       \\
ch1, ch2, ch3, ch4              & 0.74          & 0.86       \\
MIPS24                          & 0.85          & 0.86       \\
PACS100, PACS160                & 0.87          & 0.87       \\
SPIRE250, SPIRE350, SPIRE500    & 0.83          & 0.85       \\\hline    
\end{tabular}
\end{table}

\section{4-class model}
\label{app:4class}
\begin{figure}[t]
\centering
   \includegraphics[width=\hsize]{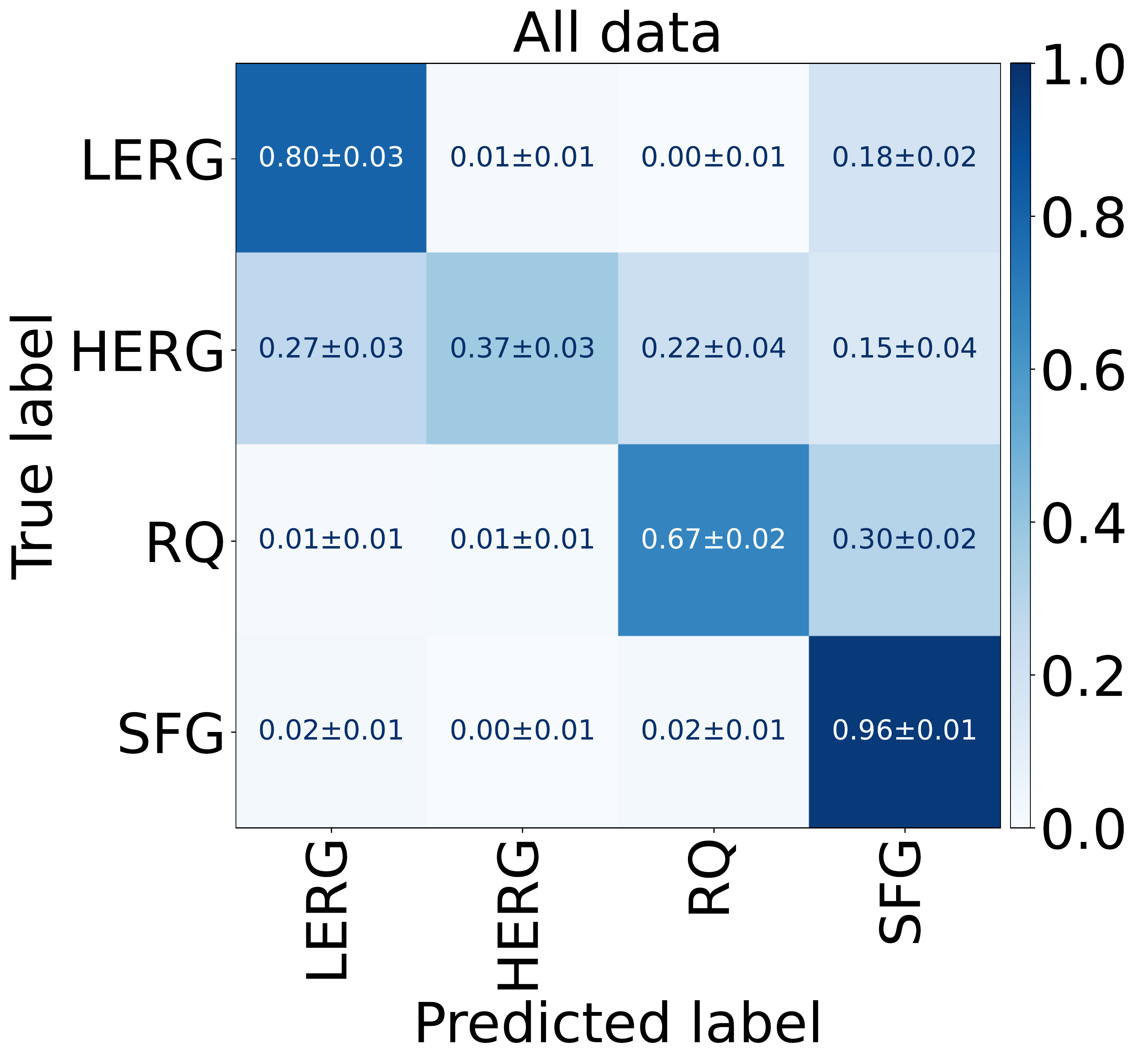}
    \caption{Confusion matrix of the cross-validated 4-class model. This confusion matrix has been created on the cross-validated testing sets. A perfect classifier has all 1's across the diagonal, and 0's everywhere else. It has been normalised over the rows, such that the diagonal represent the recalls.}
   \label{fig:4class}
\end{figure}

In addition to training a 2-class model, we also train a 4-class model. This model is trained in a similar way to the 2-class model. As for the 2-class model, we plot a confusion matrix to get an overview of the model. This confusion matrix is plotted in Fig. \ref{fig:4class}. This figure shows that the boundary between SFG and AGN is just as good as the 2-class model. However, the classification of the subclasses of AGN performs a lot worse. For the HERG class, a recall of only 38\%$\pm$6\% is achieved. For the other classes, higher scores are displayed but are still insufficient for an accurate classifier.

A large fraction of the misclassifications is AGN subclasses being misclassified as SFG. This is because SFGs are the majority class. Additionally, for the HERGs, it can be seen that they are often misclassified only in radio-loud mode (i.e., HERGs being classified as LERGs) or radiative mode (i.e., HERGs being classified as RQs).

The bad performance of the minority classes is partly due to the small size of these classes. This class imbalance could be fixed by removing a large portion of the larger classes, but this would result in very few sources remaining. The training set would then become too small to achieve a good performance. Additional data, particularly for the minority classes, is thus required. 
\end{document}